\documentclass[twocolumn]{aastex}
\usepackage{amsmath,esint}
\usepackage{url,aasmacros}
\usepackage{natbib}
\usepackage{soul, CJK}
\usepackage{multirow}

\usepackage{bm}
\usepackage{xspace}
\usepackage{color}
\usepackage{enumitem}

\usepackage[utf8]{inputenc}

\shorttitle{GROWTH on S190814bv}
\shortauthors{Andreoni \& Goldstein and the GROWTH Collaboration}

\begin{document}

\title{	
GROWTH on S190814bv: Deep Synoptic Limits on the Optical/Near-Infrared Counterpart to a Neutron Star--Black Hole Merger 
}

\author[0000-0002-8977-1498]{Igor~Andreoni}
\email{andreoni@caltech.edu}
\affiliation{California Institute of Technology, 1200 East California Blvd, MC 249-17, Pasadena, CA 91125, USA}
\author[0000-0003-3461-8661]{Daniel~A.~Goldstein}
\altaffiliation{Hubble Fellow}
%\email{danny@caltech.edu}
\affiliation{California Institute of Technology, 1200 East California Blvd, MC 249-17, Pasadena, CA 91125, USA}
\collaboration{these authors contributed equally to this work}

\author{Mansi M. Kasliwal}
\affiliation{California Institute of Technology, 1200 East California Blvd, MC 249-17, Pasadena, CA 91125, USA}

\author[0000-0002-3389-0586]{Peter E. Nugent}
\affiliation{Lawrence Berkeley National Laboratory, 1 Cyclotron Road, Berkeley, CA 94720, USA}
\affiliation{Department of Astronomy, University of California, Berkeley, CA 94720-3411, USA}

\author[0000-0001-5381-4372]{Rongpu Zhou}
\affiliation{Lawrence Berkeley National Laboratory, 1 Cyclotron Road, Berkeley, CA 94720, USA}

\author[0000-0001-8684-2222]{Jeffrey A. Newman}
\affiliation{Department of Physics and Astronomy and PITT PACC, University of Pittsburgh, PA, 15260, USA}

\author[0000-0002-8255-5127]{Mattia Bulla}
\affiliation{Nordita, KTH Royal Institute of Technology and Stockholm University, Roslagstullsbacken 23, SE-106 91 Stockholm, Sweden}
\affiliation{The Oskar Klein Centre, Department of Physics, Stockholm University, AlbaNova, SE-106 91 Stockholm, Sweden}

\author{Francois Foucart}
\affiliation{Department of Physics, University of New Hampshire, 9 Library Way, Durham NH 03824, USA}

\author{Kenta Hotokezaka}
\affiliation{Department of Astrophysical Sciences, Princeton University, Peyton Hall, Princeton, NJ 08544, USA}

\author{Ehud Nakar}
\affiliation{Department of Astrophysics, Sackler School of Physics and Astronomy, Tel Aviv University, Tel Aviv, 69978, Israel}

\author{Samaya Nissanke}
\affiliation{GRAPPA, Anton Pannekoek Institute for Astronomy and Institute of High-Energy Physics, University of Amsterdam, Science Park 904, 1098 XH Amsterdam, The Netherlands}
\affiliation{Nikhef, Science Park 105, 1098 XG Amsterdam, The Netherlands}

\author{Geert Raaijmakers}
\affiliation{GRAPPA, Anton Pannekoek Institute for Astronomy and Institute of High-Energy Physics, University of Amsterdam, Science Park 904, 1098 XH Amsterdam, The Netherlands}
\affiliation{Nikhef, Science Park 105, 1098 XG Amsterdam, The Netherlands}

\author[0000-0002-9870-5695]{Joshua S. Bloom}
\affiliation{Department of Astronomy, University of California, Berkeley, CA 94720-3411, USA}
\affiliation{Lawrence Berkeley National Laboratory, 1 Cyclotron Road, Berkeley, CA 94720, USA}

\author{Kishalay De}
\affiliation{California Institute of Technology, 1200 East California Blvd, MC 249-17, Pasadena, CA 91125, USA}

\author{Jacob E. Jencson}
\affiliation{California Institute of Technology, 1200 East California Blvd, MC 249-17, Pasadena, CA 91125, USA}
\affiliation{University of Arizona, Steward Observatory, 933 N. Cherry Avenue, Tucson, AZ 85721, USA}

\author{Charlotte Ward}
\affiliation{Department of Astronomy, University of Maryland, College Park, MD 20742, USA}

\author[0000-0002-2184-6430]{Tom{\'a}s Ahumada}
\affil{Department of Astronomy, University of Maryland, College Park, MD 20742, USA}

\author{Shreya Anand}
\affiliation{California Institute of Technology, 1200 East California Blvd, MC 249-17, Pasadena, CA 91125, USA}

\author{David A. H. Buckley}
\affiliation{South African Astronomical Observatory, PO Box 9, Observatory 7935, Cape Town, South Africa}

\author{Maria D. Caballero-Garc\'ia}
\affiliation{Astronomical Institute, Academy of Sciences of the Czech Republic, Bo\v{c}n\'{\i}~II 1401, CZ-141\,00~Prague, Czech Republic}

\author{Alberto J. Castro-Tirado}
\affiliation{Instituto de Astrof\'isica de Andaluc\'ia (IAA-CSIC), Glorieta de la Astronom\'ia s/n, E-18008, Granada, Spain}
\affiliation{ Departamento de Ingenier\'ia de Sistemas y Autom\'atica, Escuela de Ingenieros Industriales, Universidad de M\'alaga, Unidad Asociada al CSIC, C. Dr. Ortiz Ramos sn, 29071 M\'alaga, Spain}

\author[0000-0001-7983-8698]{Christopher M. Copperwheat}
\affil{Astrophysics Research Institute, Liverpool John Moores University, \\ IC2, Liverpool Science Park, 146 Brownlow Hill, Liverpool L3 5RF, UK}

\author[0000-0002-8262-2924]{Michael W. Coughlin}
\affiliation{California Institute of Technology, 1200 East California Blvd, MC 249-17, Pasadena, CA 91125, USA}

\author[0000-0003-1673-970X]{S. Bradley Cenko}
\affiliation{Astrophysics Science Division, NASA Goddard Space Flight Center, MC 661, Greenbelt, MD 20771, USA}
\affiliation{Joint Space-Science Institute, University of Maryland, College Park, MD 20742, USA}

\author{Mariusz Gromadzki}
\affiliation{Astronomical Observatory, University of Warsaw, Al. Ujazdowskie 4, 00-478 Warszawa, Poland}

\author{Youdong Hu}
\affiliation{Instituto de Astrof\'isica de Andaluc\'ia (IAA-CSIC), Glorieta de la Astronom\'ia s/n, E-18008, Granada, Spain}
\affiliation{Universidad de Granada, Facultad de Ciencias Campus Fuentenueva S/N CP 18071 Granada, Spain}

\author{Viraj R. Karambelkar }
\affiliation{California Institute of Technology, 1200 East California Blvd, MC 249-17, Pasadena, CA 91125, USA}

\author{Daniel A. Perley}
\affil{Astrophysics Research Institute, Liverpool John Moores University, \\ IC2, Liverpool Science Park, 146 Brownlow Hill, Liverpool L3 5RF, UK}

\author{Yashvi Sharma}
\affiliation{California Institute of Technology, 1200 East California Blvd, MC 249-17, Pasadena, CA 91125, USA}

\author{Azamat F. Valeev}
\affiliation{Special Astrophysical Observatory, Russian Academy of Sciences, Nizhnii Arkhyz, 369167 Russia}

\author[0000-0002-6877-7655]{David O. Cook}
\affiliation{IPAC, California Institute of Technology, 1200 E. California Blvd, Pasadena, CA 91125, USA}

\author{U. Christoffer Fremling}
\affiliation{California Institute of Technology, 1200 East California Blvd, MC 249-17, Pasadena, CA 91125, USA}

\author{Harsh Kumar}
\affiliation{Indian Institute of Technology Bombay, Powai, Mumbai 400076, India}

\author{Kirsty Taggart}
\affil{Astrophysics Research Institute, Liverpool John Moores University, \\ IC2, Liverpool Science Park, 146 Brownlow Hill, Liverpool L3 5RF, UK}

\author{Ashot Bagdasaryan}
\affiliation{California Institute of Technology, 1200 East California Blvd, MC 249-17, Pasadena, CA 91125, USA}

\author[0000-0001-5703-2108]{Jeff Cooke}
\affiliation{Australian Research Council Centre of Excellence for Gravitational Wave Discovery (OzGrav), Swinburne University of Technology, Hawthorn, VIC, 3122, Australia}
\affiliation{Centre for Astrophysics and Supercomputing, Swinburne University of Technology, Hawthorn, VIC, 3122, Australia}

\author{Aishwarya Dahiwale}
\affiliation{California Institute of Technology, 1200 East California Blvd, MC 249-17, Pasadena, CA 91125, USA}

\author[0000-0002-2376-6979]{Suhail Dhawan}
\affiliation{The Oskar Klein Centre, Department of Physics, Stockholm University, AlbaNova, SE-106 91 Stockholm, Sweden}

\author[0000-0003-0699-7019]{Dougal Dobie}
\affiliation{Sydney Institute for Astronomy, School of Physics, University of Sydney, NSW 2006, Australia}
\affiliation{CSIRO Astronomy and Space Science, P.O. Box 76, Epping, New South Wales 1710, Australia}

\author[0000-0002-1955-2230]{Pradip Gatkine}
\affil{Department of Astronomy, University of Maryland, College Park, MD 20742, USA}

\author[0000-0001-8205-2506]{V. Zach Golkhou}
\altaffiliation{Moore-Sloan, WRF Innovation in Data Science, and DIRAC Fellow}
\affiliation{DIRAC Institute, Department of Astronomy, University of Washington, 3910 15th Avenue NE, Seattle, WA 98195, USA}
\affiliation{The eScience Institute, University of Washington, Seattle, WA 98195, USA}

\author[0000-0002-4163-4996]{Ariel Goobar}
\affiliation{The Oskar Klein Centre, Department of Physics, Stockholm University, AlbaNova, SE-106 91 Stockholm, Sweden}

\author[0000-0001-9560-2220]{Andreas Guerra Chaves}
\affiliation{GRAPPA, Anton Pannekoek Institute for Astronomy and Institute of High-Energy Physics, University of Amsterdam, Science Park 904, 1098 XH Amsterdam, The Netherlands}

\author[0000-0001-9315-8437]{Matthew Hankins}
\affiliation{California Institute of Technology, 1200 East California Blvd, MC 249-17, Pasadena, CA 91125, USA}

\author[0000-0001-6295-2881]{David L. Kaplan}
\affiliation{Center for Gravitation, Cosmology and Astrophysics, Department of Physics, University of Wisconsin--Milwaukee, P.O. Box 413, Milwaukee, WI 53201, USA}

\author[0000-0002-5105-344X]{Albert K. H. Kong}
\affiliation{Institute of Astronomy, National Tsing Hua University, Hsinchu 30013, Taiwan}

\author[0000-0002-7252-3877]{Erik C. Kool}
\affiliation{The Oskar Klein Centre \& Department of Astronomy, Stockholm University, AlbaNova, SE-106 91 Stockholm, Sweden}

\author{Siddharth Mohite}
\altaffiliation{LSSTC Data Science Fellow}
\affiliation{Center for Gravitation, Cosmology and Astrophysics, Department of Physics, University of Wisconsin--Milwaukee, P.O. Box 413, Milwaukee, WI 53201, USA}

\author{Jesper Sollerman}
\affiliation{The Oskar Klein Centre \& Department of Astronomy, Stockholm University, AlbaNova, SE-106 91 Stockholm, Sweden}

\author[0000-0003-0484-3331]{Anastasios Tzanidakis}
\affiliation{California Institute of Technology, 1200 East California Blvd, MC 249-17, Pasadena, CA 91125, USA}

\author[0000-0003-2601-1472]{Sara Webb}
\affiliation{Centre for Astrophysics and Supercomputing, Swinburne University of Technology, Hawthorn, VIC, 3122, Australia}
\affiliation{Australian Research Council Centre of Excellence for Gravitational Wave Discovery (OzGrav), Swinburne University of Technology, Hawthorn, VIC, 3122, Australia}

\author[0000-0002-9870-5695]{Keming Zhang %\begin{CJK*}{UTF8}{gkai}(张可名)\end{CJK*}
}
\altaffiliation{LSSTC Data Science Fellow}
\affiliation{Department of Astronomy, University of California, Berkeley, CA 94720-3411, USA}

\begin{abstract}
On 2019 August 14, the Advanced LIGO and Virgo interferometers detected the high-significance gravitational wave (GW) signal S190814bv. The GW data indicated that the event resulted from a neutron star--black hole (NSBH) merger, or potentially a low-mass binary black hole merger. Due to the low false alarm rate and the precise localization (23 deg$^2$ at 90\%), S190814bv presented the community with the best opportunity yet to directly observe an optical/near-infrared counterpart to a NSBH merger. To search for potential counterparts, the GROWTH collaboration performed real-time image subtraction on 6 nights of public Dark Energy Camera (DECam) images acquired in the three weeks following the merger, covering $>$98\% of the localization probability. Using a worldwide network of follow-up facilities, we systematically undertook spectroscopy and imaging of optical counterpart candidates. Combining these data with a photometric redshift catalog, we ruled out each candidate as the counterpart to S190814bv and we placed deep, uniform limits on the optical emission associated with S190814bv.  For the nearest consistent GW distance, radiative transfer simulations of NSBH mergers constrain the ejecta mass of S190814bv to be $M_\mathrm{ej} < 0.04$~$M_{\odot}$ at polar viewing angles, or $M_\mathrm{ej} < 0.03$~$M_{\odot}$ if the opacity is $\kappa < 2$~cm$^2$g$^{-1}$. Assuming a tidal deformability for the neutron star at the high end of the range compatible with GW170817 results, our limits would constrain the BH spin component aligned with the orbital momentum to be $ \chi < 0.7$ for mass ratios $Q < 6$, with weaker constraints for more compact neutron stars.
We publicly release the photometry from this campaign at \href{http://www.astro.caltech.edu/~danny/static/s190814bv/}{this http url}.
\end{abstract}

\section{Introduction}
\label{sec: intro}

Mergers of binaries containing neutron stars and stellar-mass black holes (NSBH mergers) have long been theorized as potential sites of $r$-process nucleosynthesis \citep{1974ApJ...192L.145L}, that should be detectable by networks of laser interferometers as gravitational wave (GW) sources \citep{2010CQGra..27q3001A},  potentially harboring optical counterparts \citep{Metzger2012} that could be used to help constrain the equation-of-state (EOS) of dense nuclear matter \citep{2015APS..DNPAA1001G,CoDi2018b}, measure the Hubble constant H$_0$ \citep{1986Natur.323..310S}, and probe radiation hydrodynamics in asymmetric conditions and the limits of nuclear stability \citep{2016ARNPS..66...23F}.
On 2019 August 14, the LIGO and Virgo interferometers detected S190814bv, the first high-confidence GW signal associated with an NSBH merger \citep{gcnS190814bvDiscovery,gcnS190814bvUpdate}, confirming that NSBH mergers  exist and that they produce gravitational waves.

Electromagnetic emission from NSBH mergers, which is critical to achieve many of the science goals described in the previous paragraph, is currently the subject of considerable theoretical uncertainty \citep[e.g.,][]{Mingarelli2015,Hotokezaka2019, Barbieri2019}.  
At this time, it is not clear whether optical/near-infrared (NIR) counterparts to NSBH mergers  exist, and, if they do, what their properties might be.
The uncertainty in the nature of electromagnetic counterparts to NSBH mergers is driven primarily by (1) uncertainties in the optical opacity of $r$-process elements in low ionization states, which may be the dominant opacity affecting spectrum synthesis in NSBH optical counterparts (``kilonovae,'' or ``macronovae"), (2) a lack of knowledge regarding the EOS of dense nuclear matter, which directly affects the distribution of the merger ejecta and the post-merger nucleosynthesis, (3) an incomplete theoretical picture of the properties of NSBH matter outflows for all potential progenitor configurations, and (4) the complexity of the multiphysics simulations required to predict the observable properties of NSBH mergers, which at various stages must include sophisticated treatments of magnetohydrodynamics, General Relativity, neutrino transport, radiation transport, and nucleosynthesis.

The dynamics of NSBH mergers is profoundly different from the dynamics of binary neutron star (BNS) mergers (see Nakar, in preparation for a review), but their EM counterparts are expected to share some similarities. The mass ejection depends mostly on whether the tidal radius of the NS is larger or smaller than the innermost stable circular orbit (ISCO) of the BH. In the first case, a significant fraction of the NS mass is ejected. A soft EOS of the NS is the principal responsible for a large tidal radius, while the ISCO decreases with smaller BH masses and higher spin component in the binary orbital plane \citep[e.g.,][]{Foucart:2012nc, Kawaguchi2015PhRvD, Kawaguchi:2016ana}. If the tidal radius is larger than the ISCO, then as the NS approaches the tidal radius, a 0.01--0.1~M$_{\odot}$ tidal tail can be dynamically created.
These dynamical ejecta have a low electron fraction $Y_e$, favoring heavy element production via r-process. 
An accretion disk can then form around the BH with mass $\sim$0.1--0.3~M$_{\odot}$.  About $\sim 40$\% of the disk mass is ejected, constituting a secular ejecta component \citep[e.g.,][]{Siegel2018disk, Fernandez2019disk}. This component may contain polar winds with velocities 0.1--0.15~$c$ during an efficient accretion phase, followed by more isotropic, neutron-rich winds with lower velocities. Simulations suggest that $Y_e$ of the secular ejecta is higher than that of the dynamical one, especially along the polar regions, where the ejecta may be free of the high opacity lanthanides \citep{Miller2019PhRvD, Christie2019MNRAS}.  

For comparison, the GW170817 kilonova was found to have a ``red" component with mass $M \sim 0.04M_{\odot}$, likely encompassing both the tidal ejecta and the post-merger disk wind, with velocity $v\sim 0.1 c$ \citep[e.g.,][]{Kasen2017Nat}. The nature of the blue component of GW170817 is harder to explain, however polar winds from efficient accretion onto the BH formed during a NSBH merger may result in a similarly blue transient at early times.  
If the NS is disrupted within the ISCO, the mass of both the dynamical and the disk ejecta is expected to be small, $< 10^{-3}$~M$_{\odot}$, reducing the likelihood of producing any observable EM counterpart.

To help characterize the uncertain nature of electromagnetic  emission from NSBH mergers,  we present deep, synoptic, and red-sensitive limits on the optical/NIR emission from the NSBH merger S190814bv. 
We obtained the limits from public, multi-band observations of the localization region of S190814bv conducted by the Dark Energy Survey GW (DES-GW) collaboration \citep{gcn25336}, who used the Dark Energy Camera \citep[DECam,][]{Flaugher2015} to tile $>$98\% of the localization probability roughly 10 times in each of the $i$ and $z$ bands.

Section \ref{sec: S190814bv} gives an overview of the GW event and Section \ref{sec: dataset} describes the DECam follow-up.
Our analysis methods are described in Section~\ref{sec: methods} and the results of follow-up observations of candidates of interest are presented in Section~\ref{sec: results}, using a \cite{Planck2015cosmo} cosmology to compute absolute magnitudes.  
In Section \ref{subsec: galaxies}, we quantify the completeness of our galaxy catalogs. 
In Section \ref{sec: discussion}, we use the limits obtained in the preceding analysis  to constrain the ejecta mass, opacity, and viewing angle of S190814bv.  
The constraints on the ejecta mass are used to characterize the spin and the mass ratio of the progenitor binary.
We summarize our results and present concluding remarks in Section \ref{sec: conclusion}.

\section{S190814\MakeLowercase{bv}}
\label{sec: S190814bv}

 \cite{gcnS190814bvDiscovery} detected  the GW event S190814bv on 2019-08-14 21:10:39 UT, using four independent pipelines processing data from three GW interferometers (LIGO Hanford, LIGO Livingston, and Virgo) in triple coincidence. 
The false alarm rate of the event was $2 \times 10^{-33}$~Hz, or approximately one in $10^{25}$ years.
The GW event was first classified as ``Mass Gap" with $>99\%$ probability. A ``mass gap" system refers to a binary where the lighter companion has mass $3 M_\odot <M< 5 M_\odot$, and no material is expected to be ejected. 
The classification of S190814bv was revised about 12 hours later \citep{gcnS190814bvUpdate} based on new parameter estimation obtained with the LALInference offline analysis pipeline \citep{Veitch2015, Abbott2016GWa} to an ``NSBH'' event with $>$99\% probability. 
The refined analysis also indicated that there should be $< 1\%$ probability of having disrupted material surrounding the resulting compact object.
In this work, we observationally probe the presence of remnant material that could generate an optical/NIR signature (see Section \ref{sec: discussion}) and discuss the results in the context of the NSBH scenario.

LIGO/Virgo alerts with an ``NSBH'' classification refer to events in which the lighter object has  $M \leq 3 M_\odot$ and the heavier component has $M \geq 5 M_\odot$.
The maximum mass of a neutron star, according the most extreme viable EOS, is $M_\mathrm{ns,max}\approx2.8 M_\odot$ \citep{2016ARA&A..54..401O}.
It is thus possible, given the LIGO/Virgo definition of ``NSBH,'' that GW events classified as ``NSBH'' may actually be  mergers of black holes having $M \geq 5 M_\odot$ with lower-mass black holes having $M_\mathrm{ns,max} \leq M \leq 3 M_\odot$. 
As the masses of the components of S190814bv are not yet public, we cannot yet comment on  this possibility.

S190814bv was localized to 23~deg$^2$ at 90\% confidence. For comparison, the BNS merger GW170817 was localized to 28~deg$^2$ \citep{Abbott2017GW170817discovery}, then refined to 16~deg$^2$ \citep{LVC2019GWTC-1}, and the three GW event candidates including neutron stars identified during O3 before S190814bv were localized to 7461~deg$^2$ \citep[S190425z;][]{gcn24228}, 1131~deg$^2$ \citep[S190426c;][]{gcn24277}, and 1166~deg$^2$ \citep[S190510g;][]{gcnS190510grefined}, with S190510g having a significant probability of being non-astrophysical in origin.  The  precise localization of S190814bv is largely due to the fact that (1) it was detected with  three GW interferometers  and (2) it had a favorable location in the sky  with  respect to the antenna pattern of the detectors.

Despite the small localization area, the GW analysis places S190814bv at the fairly large distance of $267 \pm 52$~Mpc \citep{gcnS190814bvUpdate}. 
This corresponds to a volume of $5.26\times10^4$~Mpc$^3$ for 90\% area and $1\sigma$ distance, or a volume of $1.09\times10^5$~Mpc$^3$ for 90\% area and $2\sigma$ distance.
The distance probability distribution is broadly Gaussian (although skewed) pixel by pixel, but not over the whole map, generally. The S190814bv skymap is relatively small, so the effect is less evident than for larger skymaps, where a pixel-by-pixel approach is particularly appropriate.
We focus the analysis presented in this paper to the 2$\sigma$ volume, corresponding to the redshift range $0.037 < z < 0.081$.

\section{Dataset}
\label{sec: dataset}
\begin{figure*}
    \centering
    \includegraphics[width=0.95\textwidth]{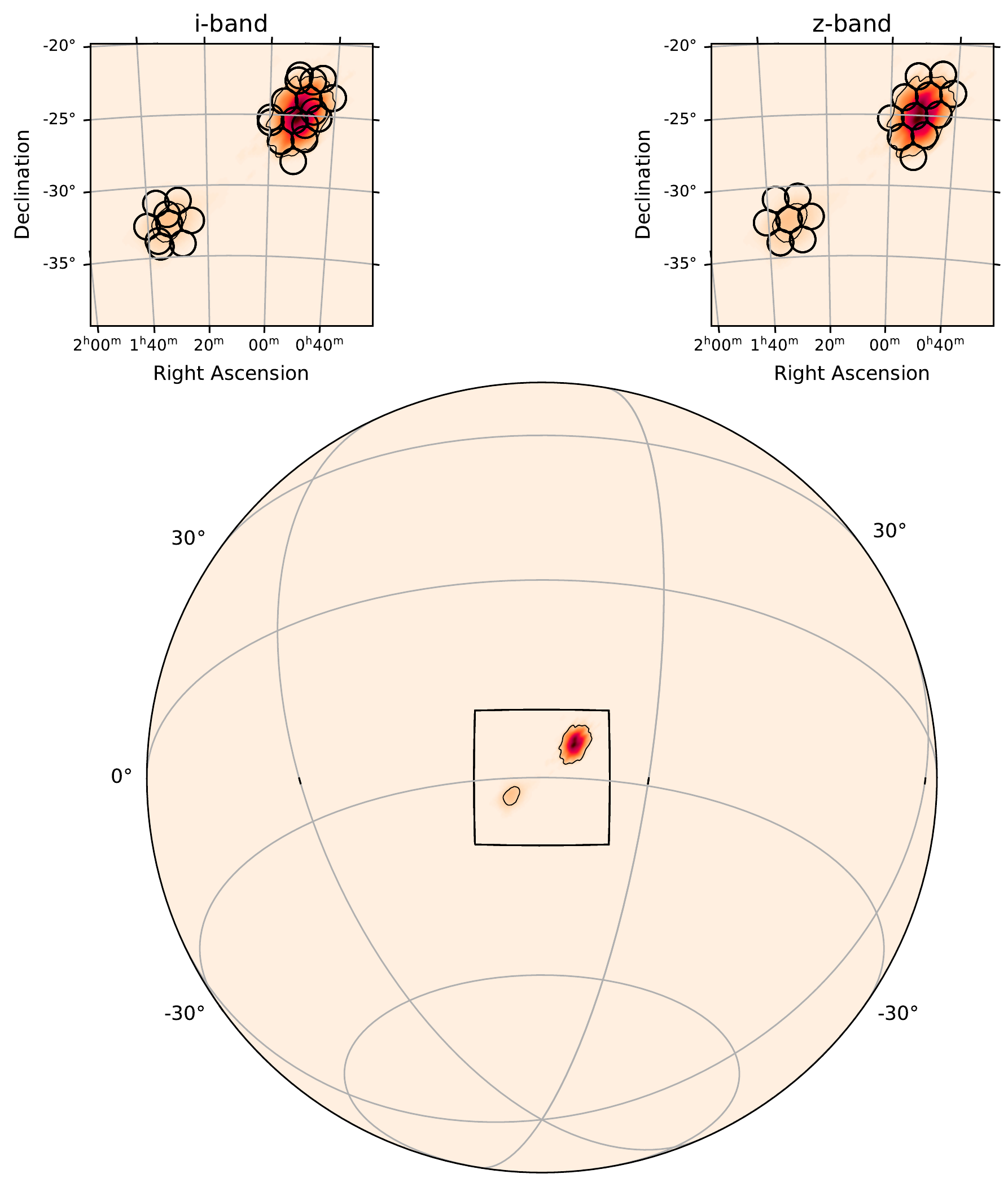}
    \caption{{\it Top row --} Locations of DECam exposures processed in this analysis (black circles) relative to the  S190814bv LALInference skymap \citep{gcnS190814bvUpdate}, with color linearly proportional to localization probability density. {\it Bottom row --} Bounding box of the top two plots (black square) relative to a global projection of the LALInference skymap.}
    \label{fig:coverage}
\end{figure*}

S190814bv was initially classified as a ``Mass Gap" event, where both the more massive object and the lighter companion are likely black holes.  Therefore, S190814bv was considered a suitable candidate for DECam follow-up under the NOAO program ID 2019B-0372 (PI Soares-Santos), which conducts observations of binary black hole (BBH) mergers, with the resulting data becoming immediately public.
The program was triggered within a few hours of the merger, before the refined classification issued by \cite{gcnS190814bvUpdate}.
The first exposure was taken roughly 7 hours after the merger at UTC 2019--08--15 06:32:43.
Data were acquired on six distinct Chilean calendar nights (2019--08--14, 2019--08--15, 2019--08--16, 2019--08--17, 2019--08--20, and 2019--08--30), lasting from 1.5 to 4.5 hours each night.
The moon and weather conditions steadily improved between the first and the last nights of the run, and the exposure times were more than twice as long in each filter at the end of the run than the beginning, resulting in a greater achieved depth.
Figure \ref{fig:coverage} shows the locations of the DECam exposures obtained during the run and processed in this analysis relative to the LALInference skymap of S190814bv.

\section{Methods}
\label{sec: methods}
\label{subsec: processing}
We processed the raw DECam data as they were taken, using the pipeline described in \cite{Goldstein2019S190426c}, now running on the Amazon Web Services Elastic Compute Cloud (EC2) for increased reliability. 
For each exposure, a \texttt{c5.18xlarge} spot EC2 instance with 72 vCPUs and 144GB of RAM was launched to astrometrically and photometrically calibrate the DECam CCD images in parallel, make references,  perform subtractions, identify candidates, filter them using \texttt{autoScan} \citep{2015AJ....150...82G}, and perform aperture photometry.
Each exposure took roughly 20 minutes to process, and the results were stored on the Amazon Simple Storage Service (S3). The median depths achieved nightly during the follow-up campaign with DECam are presented in Table~\ref{table: limiting mags}.

\subsection{Photometric redshifts}
\label{subsec: photoz}

At the distance to S190814bv  ($\sim$250 Mpc), spectroscopic redshift catalogs are largely incomplete \citep[][Cook et al, in preparation]{Dalya2018}.
We therefore relied primarily on photometric redshifts of transient host galaxies to assess whether transient candidates had distances consistent with the GW distance  of S190814bv.
We carried out an  offline analysis of the  DESI Legacy Imaging Surveys \citep{Dey2019} Data Release 8 (DR8), which includes model-based photometry from the DECam and from the {\it Wide-field Infrared Survey Explorer} \citep[{\it WISE},][]{Wright2010}, to estimate photometric redshifts for the galaxies in the S190814bv localization region.
By applying a Random Forest algorithim to the DR8 data (Zhou et al. 2019, in preparation), we generated a photometric redshift catalog for the entire DR8 footprint.

Due to its inclusion of data from Dark Energy Survey \citep[DES,][]{DES2016} observations, the catalog fully covered the S190814bv localization region. 
Catalogued sources with $m_z > 21$ were excluded because beyond that threshold the accuracy of the photometric redshifts rapidly degraded. As explained in Section~\ref{subsec: galaxies}, the impact of restricting our attention to candidates with potential host galaxies  brighter than $z = 21$ has a negligible impact on our completeness, with an expected loss in luminosity fraction of $<3\%$. 

Sources with photometric redshift uncertainties $>2\times$ the average photometric redshift uncertainty of all the sources of a similar magnitude within a 1~deg radius ($\pm 0.1$~mag) were also excluded as potential hosts.  
The distribution of photometric redshift uncertainties was estimated after rejecting stellar sources using a cut on the morphology of the best-fit light profile. We also rejected sources with {\it Gaia} parallaxes that are not compatible with $0 \pm 1.081$~mas obtained from the analysis of parallaxes measured for quasars \citep{Luri2018}.
Spectroscopic redshifts \citep[primarily from the 2dF Galaxy Redshift Survey,][]{Colless2001} were considered instead of photometric redshifts when available.

\subsection{Candidate selection}
\label{subsec: selection criteria}
We used the GROWTH Marshal \citep{Kasliwal2019marshal} to display, filter, and assess candidates detected with our image-subtraction pipeline.  During the scanning process, 519 candidates were saved that were located inside the 95\% probability area of the skymap.  

The candidates were cross-matched to known solar system objects from the IAU Minor Planet Center using the \texttt{astcheck}\footnote{\url{https://www.projectpluto.com/astcheck.htm}} utility. The cross-match radius between candidates and known solar system objects was 100\arcsec.  In addition to excluding known asteroids from our transient list, we identified elongated candidates (likely to be fast-moving uncatalogued solar system objects) by visual inspection and removed them.

The selection criteria for candidates to be reported in this work were defined as follows:

\begin{enumerate}

\item No match with moving objects reported in the IAU Minor Planet Center.

\item At least 2 detections in any filter with a time baseline of $\geq 30$ minutes to further reject fast moving objects.

\item Location within the $95\%$ probability contour of the LALInference skymap.

\item The distance to a possible host must be consistent with the distance range expected for S190814bv (accounting for $2 \times$ the standard deviation of the distance probability distribution, which translates into a redshift range of $0.037 < z < 0.081$).  The distance to host galaxies was obtained from spectroscopic or photometric redshifts available before dedicated follow-up. For the photometric redshifts, we required a host luminosity $m_z < 21$ (see Section \ref{subsec: photoz}) and, to be more conservative, we considered $2 \times$ the uncertainty on the photometric redshifts. 

\item At least 3 detections with an \texttt{autoScan} classification score $> 0.3$ to reject image-subtraction artifacts (see Section~\ref{subsec: processing}).

\item At least 10 DECam visits (including non-detections) obtained with observations on different nights, with different filters, and as a result of the dithering pattern used on individual nights in each filter. Most coordinates on the skymap had at least 20 visits (see Figure \ref{fig:coverage}).

\end{enumerate}

Candidates discovered in real time were reported to the Transient Name Server\footnote{\url{https://wis-tns.weizmann.ac.il}} (TNS).
New candidates \citep{gcn25362, gcn25391, gcn25393} and transient follow-up were reported via Gamma-ray Coordinates Network (GCN) circulars during the follow-up campaign. 
We used a radius of $20''$ to cross-match our candidates with the photometric redshift catalog (see Section~\ref{subsec: photoz}), which corresponds to a physical distance of 16~kpc at $z = 0.037$ and of 36~kpc at $z = 0.081$. 
A total of 21 candidates survived the cuts. 
The candidates passing the selection criteria are listed in Tables \ref{table: candidates classified}--\ref{table: candidates unclassified}. 

\begin{table*}[!ht]
    \centering
    \begin{tabular}{ccccccccc}
    \hline \hline
    Average Date  & $\Delta t$ & $m_{lim,i}$ & $m_{lim,z}$ & $m_{lim,i}$ & $m_{lim,z}$ & $P_{enc}$ & $P_{enc}$ & $P_{enc}$\\
    (UT) & (days) & $5\sigma$-phot  & $5\sigma$-phot & detection limit & detection limit & ($i$) & ($z$) & ($i+z$)  \\
    \hline
    2019-08-15 08:18 & 0.46 & 21.1 & 20.9 & 20.4 & 20.3 & 92\% & 94\% & 94\% \\
    2019-08-16 07:57 & 1.45 & 21.8 & 22.0 & 21.0 & 21.1 & 97\% & 97\% & 98\% \\
    2019-08-17 06:59 & 2.41 & 22.3 & 22.3 & 21.3 & 21.4 & 97\% & 97\% & 98\% \\
    2019-08-18 07:32 & 3.43 & 22.9 & 22.9 & 22.1 & 22.3 & 97\% & 97\% & 98\% \\
    2019-08-21 06:21 & 6.38 & 23.4 & 23.2 & 22.8 & 22.6 & 93\% & 93\% & 94\% \\
    2019-08-31 06:11 & 16.37 & 24.2 & \nodata & 23.4 & \nodata & 63\% & \nodata & 63\% \\
    \hline
    \end{tabular}
    \caption{Median depth achieved  during the follow-up of S190814bv. The dates correspond to the central time between the first and the last epoch acquired on each observing night, and $\Delta t$ indicates the time lag from the merger time \citep{gcnS190814bvDiscovery}. The photometric depth corresponds to 5$\sigma$ photometric magnitude limits (column 3 and 4) and detection depth indicates the detection limit of the image-subtraction pipeline. All magnitudes are calibrated to the AB system. The last three columns present the integrated probability of the S190814bv LALInference skymap observed on each observing night, with the last column considering the observations in either $i$ or $z$ filters. }
    \label{table: limiting mags}
\end{table*}

\subsection{Candidate follow-up methods}
\label{subsec: follow-up methods}

The spectroscopic results presented in this paper include data obtained using Near Infrared Echellete Spectrometer (NIRES) and the Low Resolution Imaging Spectrometer \citep[LRIS,][]{Oke1995} at W.~M. Keck Observatory. The NIRES data were reduced using the \texttt{Spextool} code \citep{Cushing2004} adapted for NIRES. The LRIS data were processed using \texttt{lpipe}, the fully automated reduction pipeline for longslit spectroscopy described in \cite{Perley2019lpipe}.
We observed three potential candidates with
the 10.4m Gran Telescopio de Canarias (GTC, PI A. Castro-Tirado), located at the observatory of Roque de los Muchachos in La Palma (Canary Islands, Spain), equipped with the Optical System for
Imaging and low-intermediate-Resolution Integrated Spectroscopy
\citep[OSIRIS,][]{Cepa2000}.
GTC/OSIRIS spectra for the three targets were obtained
either with the R1000B or with the R1000R grisms and a 1~arcsec slit
covering the 3,700{\AA}--7,500{\AA} or 5,100{\AA}--10,000{\AA} range. The slit was placed in order to cover the candidate location and the host galaxy centre. Data were reduced and calibrated using standard routines.
Optical images in the $r$-band filter were also taken for the candidates with GTC. Photometric zero points and astrometric calibration were computed using the Pan-STARRS catalogue \citep{Chambers2016arXiv}. We then performed point spread function (PSF) matching photometry of the targets.
Spectroscopy of one candidate of interest was also obtained with 10m Southern African Large Telescope \citep[SALT,][PI Buckley]{Buckley2006} equipped with the Robert Stobie Spectrograph \citep[RSS,][]{Burgh2003,Kobulnicky2003}.
The primary data reduction of the SALT/RSS spectrum was done using the \texttt{PySALT} package \citep{Crawford2010}, which accounts for basic CCD characteristics (e.g., cross-talk, bias and gain correction) and cosmic ray removal. Standard \texttt{IRAF/Pyraf} routines were then used to undertake wavelength and relative flux calibrations. Due to the design of SALT, which has a changing
field-dependent entrance pupil, spectrophotometric standard observations can only provide relative fluxes \citep{Buckley2018}.

The photometric evolution of the most promising candidates was monitored using the optical imaging component of the Infrared-Optical suite of instruments (IO:O) on the 2m Liverpool Telescope \citep[LT,][]{Steele2004} at Observatorio del Roque de los Muchachos. All images were processed with the LT IO:O pipeline and image subtraction was performed automatically using Pan-STARRS \citep{Chambers2016arXiv} imaging as a reference, using the methods described in \cite{Fremling+2016}.
Optical photometric follow-up data were also acquired using the Las Cumbres Observatory (LCO) telescope network under proposal ID 2019B-0244 (PI Coughlin).
The LCO photometry was measured after subtracting reference images from the Legacy Surveys archive using the \texttt{HOTPANTS} package \citep{Becker2015}.  At infrared wavelengths we obtained photometry using the Wide-field Infrared Camera \citep[WIRC,][]{Wilson2003} on the Palomar 200-inch Hale telescope (P200). The P200/WIRC data were reduced using a  reduction pipeline developed by members of our team (De et al., in preparation).

\section{Results}
\label{sec: results}

In this section, we describe the follow-up observations that were conducted to characterize each of the 21 objects that we selected as candidate counterparts to S190814bv using the methods described in Section~\ref{sec: methods}. In addition, we discuss a selection of candidates that did not pass our selection criteria, but that were reported and extensively followed up in the first three weeks after S190814bv. Most of the objects presented individually were spectroscopically classified.

\begin{figure*}[t]
    \centering
        \includegraphics[width=\textwidth]{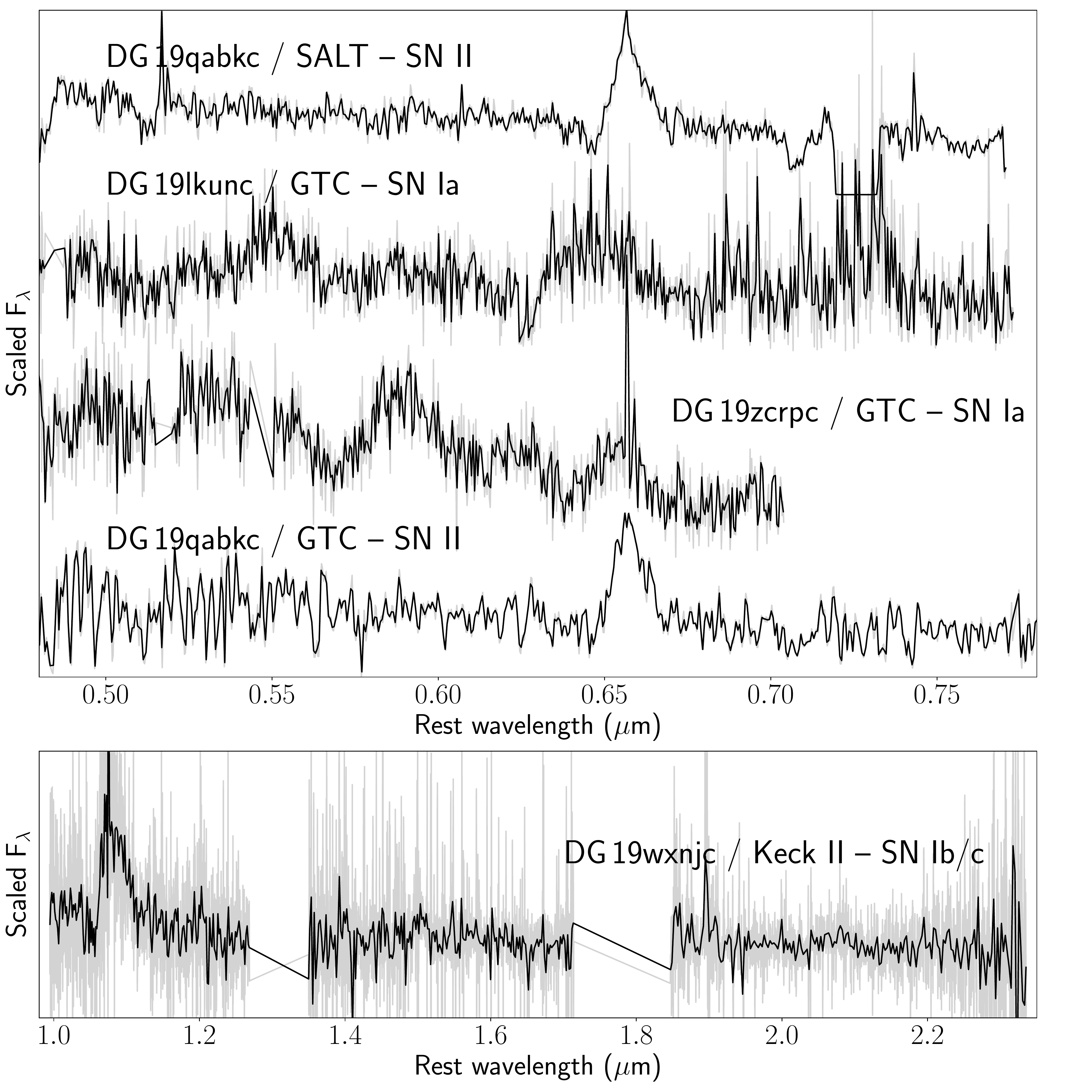}
    \caption{{\it Top panel --} Optical spectroscopic follow-up of candidates found in the localization region. The black lines correspond to binned versions of the unbinned reduced spectra shown in gray. GTC / OSIRIS and SALT spectra of DG\,19qabkc show a strong P-Cygni H$\alpha$ line suggesting a Type II supernova at $z = 0.08$. GTC / OSIRIS spectra of DG19lkunc and DG19zcrpc are consistent with SNe\,Ia at $z = 0.21$ and $z=0.08$ respectively. {\it Lower panel --} Near-infrared spectrum of DG19wxnjc obtained with Keck II / NIRES. The spectrum shows a prominent P-Cygni feature at $\approx 1.08 \mu$m, consistent with He I with an absorption velocity of 7000 km s$^{-1}$, classifying this source as a Type Ib/c supernova. These classifications rule out associations of these sources to S190814bv.}
    \label{fig:spectra}
\end{figure*}

\begin{figure*}[t]
    \centering
    \includegraphics[width=1\textwidth]{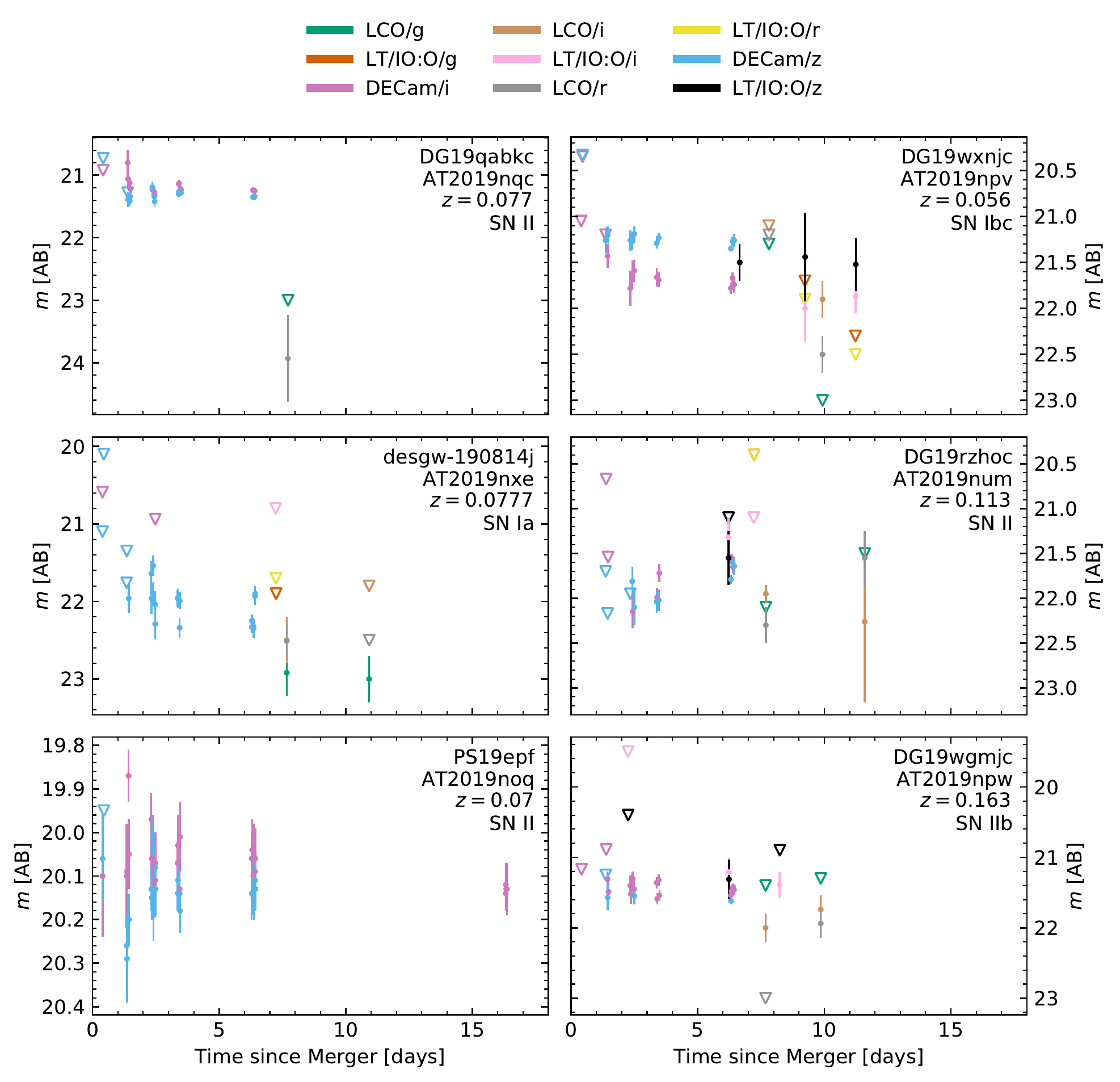}
    \caption{Light curves of the first six candidates presented in Table~\ref{table: candidates classified}. The LCO and LT photometry upper limits are quoted to 3$\sigma$, while DECam upper limits are quoted to $5\sigma$. P200/WIRC photometry was not plotted because it is host contaminated.
    Photometry for all candidates in Tables \ref{table: candidates classified} and \ref{table: candidates unclassified} is available in machine-readable form online (data behind the figure). We note that absolute magnitudes were not K corrected and must be considered to be indicative values.} 
    \label{fig: lc 1}
\end{figure*}

\paragraph{DG19qabkc/AT2019nqc} 
The candidate was first reported in \cite{gcn25362} and appeared to be $\sim 2''$ offset from its host galaxy.  Although no spectroscopic redshift was available, the photometric redshift placed the host in the correct distance range \citep{gcn25391}.  The candidate was photometrically confirmed in the optical \citep{gcn25373, gcn25374, gcn25397} and we detected the transient in the near infrared at magnitude $J \sim 21.4 \pm 0.2$ on 2019-08-18 using P200/WIRC \citep{gcn25396}. A flux upper limit of $F < 2.9 \times 10^{-12}$~erg~cm$^{-2}$~s$^{-1}$ \citep{gcn25400} was placed using data acquired with the X--ray Telescope \citep[XRT,][]{Burrows2005} on the space-based {\it Neil Gehrels Swift Observatory}, hereafter referred to as {\it Swift}.  
We observed DG19qabkc/AT2019nqc with SALT/RSS starting on 2019-08-23 22:46:10 and two consecutive 1200~s exposures were obtained using the PG300 transmission grating, which covered the spectral region 3300--9800{\AA}. The seeing was $\sim 1.7''$ and a $1.5''$ slit was used, giving an average resolving power of $\sim 370$, or a resolution of $\sim 18${\AA}. 
A strong broad H$\alpha$ line with a P-Cygni profile dominates the spectrum, with a weak H$\beta$ line in absorption, consistent with a redshift of $z=0.077$.  A good match was obtained using SNID \citep{Blondin2007} using SN2005cs, a SN type II, 14 days after maximum \citep[Figure~\ref{fig:spectra}, see also][]{gcn25481}.
Our GTC/OSIRIS spectroscopic observations confirmed DG19qabkc/AT2019nqc to be a SN II at $z = 0.078 \pm 0.001$ \citep{gcn25571}.
We extensively monitored the transient with LCO and LT imaging.  The photometry that we obtained (data behind Figure~\ref{fig: lc 1}) confirms a slow evolution compatible with supernova behavior.

\paragraph{DG19wxnjc/AT2019npv}

When the candidate was discovered \citep{gcn25393} it appeared to be offset from its host galaxy, the  photometric redshift of which ($z=0.072 \pm 0.056$) was compatible with the expected distance to S190814bv. The redshift of the host was spectroscopically measured to be $z = 0.056$ \citep{gcn25454}. The candidate was later confirmed in the optical and near-infrared \citep{gcn25398, gcn25455, gcn25457, gcn25474,gcn25485}, but no X--ray counterpart was detected with {\it Swift}/XRT \citep[$F < 3.8 \times 10^{-12}$~erg~cm$^{-2}$~s$^{-1}$,][]{gcn25400} and no radio counterpart was detected with the Australian Square Kilometre Array Pathfinder (ASKAP) \citep[$S_{\rm 943 MHz} < 75\,\mu$Jy;][]{gcn25472} and Karl G. Jansky Very Large Array (VLA) \citep[$S_{\rm 6GHz} < 12\,\mu$Jy;][]{gcn25480}. \cite{gcn25468} reported a possible archival detection in DES data, questioning the transient nature of DG19wxnjc/AT2019npv. \cite{gcn25458} produced precise photometry obtained with nightly stacks of DECam data, indicating the transient to be reddening at a rate $\Delta (i-z) \sim 0.05$~mag~day$^{-1}$. 
%\cite{gcn25460} Christoffer's photometry
The transient was also monitored photometrically with LCO and LT \citep[see][and data behind Figure~\ref{fig: lc 1}]{gcn25477}, which produced detections in the $r,i$ and $z$ bands and a marginal detection $g \gtrsim 23.0$ on 2019-08-24 with LCO, further indicating the transient to be (or to have become) red in color. %\textcolor{red}{[Comment on the photometric evolution]} 
We obtained one epoch of P200/WIRC imaging of the transient in $J$ band, and did not detect the source to a 5$\sigma$ limit of 21.4 AB mag, although we caution that the photometry is contaminated by host galaxy light.

We obtained one NIR spectrum of DG19wxnjc with NIRES on the Keck II telescope on 2019 August 24 \citep{gcn25461, gcn25478}. We acquired two sets of dithered ABBA exposures on the transient location for a total exposure time of 40 mins. 
The telluric standard HIP 7202 was used for flux calibration. The reduced and stacked spectra showed a largely featureless continuum between 1.0 and 2.5 $\mu$m (Figure~\ref{fig:spectra}) along with a prominent P-Cygni profile near 1.08 $\mu$m with an absorption velocity of $\approx 7000$ km s$^{-1}$. This feature is consistent with He I at the redshift of the host galaxy, in addition to a weak hint for another He I feature at 2.05 $\mu$m, confirming the classification of this source as a Type Ib/c supernova and unrelated to S190814bv. \cite{gcn25483} confirmed the SN Ib classification using the IMACS optical spectrograph on the Magellan telescope.
%DES archival identification \cite{gcn25468}

\paragraph{desgw-190814j/AT2019nxe}
The candidate was announced by \cite{gcn25425} and was independently detected with our image-subtraction pipeline on multiple $z$-band epochs with internal name DG19zcrpc. The photometric redshift of the host is $z = 0.106 \pm 0.035$.
LCO photometry (data behind Figure~\ref{fig: lc 1}) suggests no significant $g$-band evolution between 2019-08-22 and 2019-08-25 and color $r - i \simeq 0$ on 2019-08-22.
The transient was observed with GTC in imaging and spectroscopy mode on 2019-08-23. The GTC/OSIRIS spectrum of desgw-190814j/AT2019nxe is compatible with a SN Ia at redshift $z = 0.0777 \pm 0.0005$ \citep{gcn25543}. 
%In final list of DES candidates. \cite{gcn25486}

\paragraph{DG19rzhoc/AT2019num} We identified this candidate in DECam data \citep{gcn25393} and it was independently confirmed in the same dataset \citep{gcn25398}, in images taken with the Reionization and Transients Infrared Camera (RATIR\footnote{\url{www.ratir.org}}) on the 1.5m Harold Johnson Telescope at the Observatorio Astronomico Nacional on Sierra San Pedro Martir \citep{gcn25416}, and in VLT Survey Telescope (VST) images \citep{gcn25748}. We performed photometric follow-up with LCO and LT (data behind Figure~\ref{fig: lc 1}) which revealed the transient to be slowly evolving on day timescales. A {\it Swift}/XRT upper limit was placed at $F < 4.1 \times 10^{-12}$~erg~cm$^{-2}$~s$^{-1}$ \citep{gcn25400}. We obtained one epoch of P200/WIRC imaging of the source in $J$ band and did not detect the source to a 5$\sigma$ limit of 21.4 AB mag, although we caution that the transient location is contaminated heavily with host galaxy light.
%\cite{gcn25460} Christoffer's photometry
DG19rzhoc/AT2019num was spectroscopically classified as a Type II SN at redshift $z=0.113$ using the Goodman High Throughput Spectrograph (GHTS) on the 4.1m Southern Astrophysical Research (SOAR) telescope \citep{gcn25484}. 
%TAROT non-detection \cite{gcn25599} 

\paragraph{PS19epf/AT2019noq} The candidate was identified with the Pan-STARRS1 telescope and reported on 2019-08-15 \citep{gcn25356}. We independently detected PS19epf/AT2019noq in DECam data starting on 2019-08-15 06:44:29 with internal name DG19lsugc. The transient was classified as SN II at redshift $z=0.07$ using SOAR/GHTS \citep{gcn25423}. A pre-detection of the transient in ZTF data 2 weeks before the GW event further excluded its association with S190814bv.

\paragraph{DG19wgmjc/AT2019npw} This candidate was discovered in DECam data \citep{gcn25362} and flagged as a high priority target because of the photometric redshift of the putative host $z = 0.140 \pm 0.054$ being compatible with the distance of S190814bv \citep{gcn25391}.
The transient was confirmed with optical observations with other telescopes such as the Discovery Channel Telescope \citep[DCT;][]{gcn25374, gcn25397}, VST \citep{gcn25748}, and with our P200/WIRC imaging observations \citep{gcn25396}. Photometric measurements on DECam data indicated no rapid optical evolution \citep{gcn25460}. A {\it Swift}/XRT upper limit was placed at $F < 4.3 \times 10^{-12}$~erg~cm$^{-2}$~s$^{-1}$ \citep{gcn25400}. DG19wgmjc/AT2019npw was eventually classified as a Type IIb SN at redshift $z = 0.163$ using SOAR/GHTS \citep{gcn25484}.

\paragraph{DG19sbzkc/AT2019ntr} We initially identified this candidate in DECam data \citep{gcn25393} %Correction to another candidate's info \cite{gcn25394} 
and the detection was confirmed using RATIR \citep{gcn25416} and VST \citep{gcn25748}. We note that this transient did not pass the stricter selection criteria adopted in this work (see Section~\ref{subsec: selection criteria}) because its location was visited 5 times, less than the 10-visit threshold that we imposed. DG19sbzkc/AT2019ntr was spectroscopically classified as a SN II at redshift $z = 0.2$ using SOAR/GHTS \citep{gcn25540}.

\paragraph{desgw-190814q/AT2019obc} The candidate was found and announced by the DESGW team \citep{gcn25438} and we independently detected it with our automatic pipeline from 2019-08-16 05:56:59 with internal name DG19lkunc. Our DECam photometry using Pan-STARRS1 templates was consistent with a flat evolution until 2019-08-21 \citep{gcn25460}. We acquired P200/WIRC near-infrared imaging in Ks band on MJD 58\,717.492 and the transient was not detected down to a 5$\sigma$ limit of $Ks > 20.72$ AB magnitude. desgw-190814q/AT2019obc was classified as a SN Ia few days past its peak at redshift $z = 0.216 \pm 0.005$ using GTC/OSIRIS \citep{gcn25543}.

\paragraph{ZTF19abkfmjp/SN2019mbq} The transient was discovered with ZTF on 2019-07-30 \citep{2019TNSTR1370....1N}, before S190814bv, and it was classified as a SN II at redshift $z = 0.104 \pm 0.013$ with the SED Machine \citep{2018PASP..130c5003B} on the 60-inch telescope at Palomar Observatory. We automatically found the transient (dubbed DG19fcmgc) in DECam data and it was also reported by two other groups via GCN \citep{gcn25336,gcn25748}. Given the pre-detection with ZTF and the SN classification, ZTF19abkfmjp/SN2019mbq cannot be associated with S190814bv.

\paragraph{DG19gxuqc/AT2019paa} We obtained a spectrum of this nuclear candidate with Keck/LRIS. The spectrum was host-dominated, with common emission lines from the galaxy that allowed us to place the host at redshift $z=0.191$, beyond the acceptable distance range for S190814bv. \\

%###############

All candidates in Table~\ref{table: candidates unclassified} are ruled out based on their photometric evolution being slower than 0.1 mag day$^{-1}$. 
This limit was adopted based on the photometric evolution of GW170817, the best-studied kilonova to date. 
GW170817 faded  faster (almost 2 mag in $g$ in 24 hours) and reddened faster (from $g-z=-0.3$ to $+1.3$ in 24 hours) than any known or theorized transient \citep[e.g.,][]{Cowperthwaite2017,Kilpatrick2017Sci, Drout2017Sci,Kasliwal2017}. 
Theoretical models \citep[e.g.,][]{Tanaka2018, Bulla2019, Hotokezaka2019} also suggest that kilonovae arising from mergers with at least one NS are rapidly evolving transients.
Requiring a photometric evolution faster than  0.1 mag/day to be considered a counterpart candidate is thus conservative, corresponding to evolution more than an order of magnitude slower than GW170817.

Other transients were published via GCN circulars, but were not reported in Table~\ref{table: candidates classified}--\ref{table: candidates unclassified} because they did not pass our selection criteria or they were not detected with our pipeline. \cite{gcn25486} published a complete list of candidates that they identified in DECam data in the first five nights of observations, including candidates identified in deep nightly stacks.  In addition to a number of candidates already discussed in this section, we can associate only two more candidates to galaxies with photometric redshifts compatible with the distance to S190814bv.  In particular, DG19zujoc/2019oac is located outside the 95\% probability area of the LALInference skymap and desgw-190814z/AT2019omx did not have enough visits to pass our selection criteria (Section~\ref{subsec: selection criteria}).  In the available epochs, we do not measure any significant variability. Its photometric redshift of $z = 0.21 \pm 0.07$ passed our selection because we considered twice the uncertainties on photometric redshifts, however its large value suggests that the host galaxy is well beyond the distance range of interest. 

Three DECam candidates DG19zoonc/AT2019nyy, DG19gyvx/AT2019thm, and DG19ggesc/AT2019paw lie within a $20''$ radius from galaxies with photometric redshift compatible with S190814bv, however underlying galaxies at larger redshifts are most likely their host. DG19ggesc/AT2019paw is also coincident with a red stellar source detected with VISTA \citep{Greggio2014}. Therefore we exclude that DG19zoonc/AT2019nyy, DG19gyvx/AT2019thm, or DG19ggesc/AT2019paw are associated with S190814bv.

Two objects labelled desgw-190814a/AT2019nmd and desgw-190814b/AT2019nme were reported as transients possibly associated with S190814bv \citep{gcn25336}. These candidates were followed up with several telescopes whose observations resulted in non-detections \citep{gcn25346,gcn25356,gcn25392,gcn25400,gcn25599}. Querying the IAU Minor Planet Center, we found that desgw-190814a/AT2019nmd is consistent with the known asteroid (297025) 2010 GA33 \citep{gcn25348}. Inspection of DECam images allowed us to show that desgw-190814b/AT2019nme is a Solar System fast moving object \citep{gcn25355} absent from the Minor Planet Center database. In conclusion, both desgw-190814a/AT2019nmd and desgw-190814b/AT2019nme were moving objects unrelated to S190814bv.

The transient labelled desgw-190814d/AT2019nqr was first reported by \citep{gcn25373}. The candidate was detected twice with our automated pipeline (internal name DG19pihic), but the two detections occurred only 2.2 minutes apart on 2019-08-16, too close in time to pass our selection criteria of $>30$ minutes between the first and last detection. desgw-190814d/AT2019nqr was later classified as a SN IIb using SOAR/GHTS \citep{gcn25379}.

The candidate desgw-190814c/AT2019nqq \citep{gcn25373,gcn25379,gcn25391,gcn25396,gcn25398,gcn25401,gcn25416} was automatically detected with our pipeline (dubbed DG19kxqic), but it was not included in Table~\ref{table: candidates classified} because it lies outside the LALInference 95\% probability area of S190814bv. desgw-190814c/AT2019nqq was classified as SN II at $z = 0.071 \pm 0.001$ using GTC/OSIRIS \citep{gcn25419}. 

Follow-up observations were performed also for the candidates named desgw-190814f/AT2019nte \citep{gcn25398} and desgw-190814r/AT2019odc \citep{gcn25486}. The redshifts of their putative host galaxies were fixed to $z = 0.054 \pm 0.001$ for desgw-190814r/AT2019odc \citep[as part of our GTC/OSIRIS observations,][]{gcn25588} and $z=0.0702$ from the 2dF Galaxy Redshift Survey. However, both candidates did not pass the quality and reliability checks in our pipeline, in agreement with the non-detection of transient signatures in the spectra.

Four additional candidates that were detected with our pipeline were spectroscopically classified as SNe. The photometric redshift of their putative hosts placed them beyond the distance range for S190814bv. In particular, the spectrum of DG19rtekc/AT2019ntn \citep{gcn25393} obtained with SOAR/GHTS is consistent with a SN Ia-CSM or a SN IIn at $z=0.1$ \citep{gcn25423}; 
the GTC/OSIRIS spectrum of desgw-190814v/AT2019omt \citep{gcn25486} is consistent with a SN II at $z=0.1564 \pm 0.0005$ \citep{gcn25588}; 
DG19gcwjc/AT2019ntp \citep{gcn25393} was classified as a broad-line SN Ic with SOAR/GHTS \citep{gcn25596}; finally, the candidate desgw-190814e/AT2019nqs \citep{gcn25373, gcn25374} was classified as a Type Ia or Type Ibc SN at redshift $z = 0.1263$ using the X-shooter instrument on the Very Large Telescope \citep[VLT;][]{gcn25384}.

In summary, none of the transients unveiled during this follow-up campaign appears to be a viable electromagnetic counterpart to the NSBH merger S190814bv.

\startlongtable
\begin{deluxetable*}{cccccccp{44mm}}
    \tabletypesize{\footnotesize}
    \tablecaption{Subset of candidates discovered or independently detected by the DECam-GROWTH team during the follow-up of S190814bv that were spectroscopically classified. None of them is a viable optical counterpart to S190814bv. The reported candidates passed the selection criteria described in Section~\ref{subsec: selection criteria}. Specifically, they lie within the 95\% probability region of the LALInference skymap and are within 20$''$ from galaxies whose redshifts (2$\sigma$ uncertainty) are compatible with the LIGO/Virgo distance (2$\sigma$). All the transients reported in this table were detected using the image-subtraction pipeline described in Section~\ref{subsec: processing}. [$*$] We note that DG19sbzkc was observed with $<10$ visits and was added to this table for completeness.   
    \label{table: candidates classified}}
    %\title{Magnitude limits (AB)}
    \tablehead{\colhead{Name} & \colhead{IAU Name} & \colhead{RA [deg]} & \colhead{Decl. [deg]} &  \colhead{Offset $[\prime\prime]$}  & \colhead{spec-$z$} & \colhead{Classification} &  \colhead{References} }
    \startdata
DG19qabkc & AT2019nqc & $22.265296$ & $-32.705155$ & 2.2 & 0.078 $\pm$0.001 & SN II & \cite{gcn25362,gcn25481,gcn25571,gcn25373,gcn25374,gcn25391,gcn25396,gcn25397,gcn25400,gcn25401}\\
\multirow{1}{*}{DG19wxnjc} & AT2019npv & $13.384653$ & $-23.832918$  & 2.1 & 0.056 & SN Ibc & \cite{gcn25393,gcn25478,gcn25483,gcn25398,gcn25400,gcn25454,gcn25455,gcn25457,gcn25458,gcn25460,gcn25461,gcn25468,gcn25472,gcn25474,gcn25477,gcn25480,gcn25483,gcn25485}\\ 
%DG19joaqc & AT2019nup & -- & -- & -- & 0.28 & 1.182 & 0.065 $\pm$0.012 & -- & -- -- --\\ 
desgw-190814j & AT2019nxe & $11.570153$ & $-24.372559$ & 1.9 & 0.0777 & SN Ia & \cite{gcn25425,gcn25543,gcn25486}\\ 
DG19rzhoc&AT2019num & $13.881677$& $-22.969021$& 1.9 & 0.113 &SN II& \cite{gcn25393,gcn25398,gcn25400,gcn25416,gcn25460,gcn25484,gcn25599,gcn25748} \\
PS19epf & AT2019noq & $12.199507$ & $-25.306523$ & 5.0 & 0.07 & SN II & \cite{gcn25356,gcn25423} \\
DG19wgmjc & AT2019npw & 13.968326 & $-25.783301$  & 1.3 &0.163 & SN IIb & \cite{gcn25362,gcn25374,gcn25391,gcn25396,gcn25397,gcn25398,gcn25400,gcn25460,gcn25484,gcn25599,gcn25748} \\ 
DG19sbzkc* & AT2019ntr & 15.007883 & $-26.714390$ & 11.6&0.2 & SN II &  \cite{gcn25393,gcn25394,gcn25416,gcn25486,gcn25540,gcn25599,gcn25748} \\ 
desgw-190814q & AT2019obc & 14.566689 & $-24.139699$ & 1.4 & 0.216 $\pm$0.005 & SN Ia & \cite{gcn25438,gcn25460,gcn25486,gcn25543} \\ 
ZTF19abkfmjp & SN2019mbq & 10.835364 & $-25.883974$ & 1.0 &$0.104 \pm 0.013$ & SN II &\cite{gcn25336,gcn25748} \\ 
 \enddata

\end{deluxetable*}

\startlongtable
\begin{deluxetable*}{ccccrccrrr}
    \tablecaption{Additional candidates discovered during the follow-up of S190814bv whose host galaxy redshift is compatible with the LIGO/Virgo distance ($2\sigma$). These candidates are ruled out based on photometric evolution. DG19tedsc was detected for the first time on 2019-08-21 in $i$ band, which suggests a slow evolution. The reported candidates passed the selection criteria described in Section~\ref{subsec: selection criteria}.     \label{table: candidates unclassified}}
    \tabletypesize{\footnotesize}
    \tablehead{\colhead{Name}  & \colhead{IAU Name} & \colhead{RA [deg]} & \colhead{Decl. [deg]}  & \colhead{Offset $[\prime\prime]$} &     \colhead{$z_\mathrm{phot}$} & \colhead{$\sigma_z$} & \colhead{$\langle m_i - m_z\rangle$} & \colhead{$\langle \dot{m}_i
    \rangle$ [mag/day]}  & \colhead{$\langle \dot{m}_z \rangle$ [mag/day]}}
    \startdata
  DG19aferc & AT2019tig & $14.517916$ & $-26.083013$ &   $0.08$ &  $0.074$ &  $0.03$ &  \nodata &   $-0.01$ &  \nodata \\
  DG19gxuqc & AT2019paa &  $13.807414$ & $-24.119017$ &   $0.40$ &  $0.116$ &  $0.06$ &     $0.01$ &    $-0.09$ &     $0.09$ \\
  PS19ekf$^{a}$ & AT2019nbp &  $11.739110$ & $-24.361751$ &   $0.42$ &  $0.102$ &  $0.01$ &   $-0.18$ &    $0.01$ &    $0.03$ \\
  DG19hqhjc & AT2019nuj & $12.257212$ & $-23.234668$ &   $0.25$ &  $0.074$ &  $0.12$ &     $0.01$ &    $-0.01$ &    $0.00$ \\
  DG19ilqnc & AT2019tih & $11.861086$ & $-27.600835$ &  $14.41$ &  $0.217$ &  $0.08$ &  \nodata &    $0.03$ &  \nodata \\
  DG19kpykc & AT2019nul & $13.818560$ & $-26.943068$ &   $0.44$ &  $0.095$ &  $0.01$ &   $-0.23$ &   $-0.07$ &    $-0.09$ \\
  DG19tedsc & AT2019tii &  $12.396721$ & $-27.035924$ &   $0.21$ &  $0.055$ &  $0.03$ &  \nodata &  \nodata &  \nodata \\
  DG19wynuc & AT2019tij & $12.232094$ & $-22.393476$ &   $0.32$ &  $0.157$ &  $0.11$ &    $0.43$ &   $-0.06$ &    $-0.02$ \\
  DG19bpkf & AT2019tiw &  $15.022907$ & $-24.950557$ &   $0.70$ &  $0.218$ &  $0.07$ &  \nodata &   $0.01$ &  \nodata \\
  DG19bown  & AT2019tix  &$12.190325$ & $-24.647386$ &   $4.27$ &  $0.190$ &  $0.07$ &    $-0.14$ &    $0.02$ &    $0.01$ \\
  DG19ggesc & AT2019paw &$12.142854$ & $-25.090528$ &  $19.47$ &  $0.285$ &  $0.13$ &     $0.23$ &    $0.00$ &     $0.03$ \\
  DG19zoonc & AT2019nyy & $12.069377$ & $-26.640810$ &  $11.52$ &  $0.212$ &  $0.07$ &  \nodata &  \nodata &   $-0.01$ \\
  DG19gyvx  & AT2019thm  & $11.985939$ & $-26.900779$ &  $18.73$ &  $0.233$ &  $0.09$ &  \nodata &  \nodata &  0.03 \\
    \enddata
    \vspace{0.1 in}
    [$a$] DG19hcsgc, with Pan-STARRS1 pre-discovery on 2019-08-09  
 \end{deluxetable*}

\section{Galaxy Catalog Completeness}
\label{subsec: galaxies}

The completeness of a synoptic follow-up campaign such as the one conducted with DECam for S190814bv is mainly limited by the area  covered and the efficiency of the transient detection pipeline.  Once these two quantities are set, the ability to detect an EM counterpart becomes flux limited. Given a mean detection limit of 21.7 mag (Table~\ref{table: limiting mags}), we were able to find transients with absolute magnitude $M \leq -14.4$ at a distance $D = 163$~Mpc, $M \leq -15.4$ at $D = 267$~Mpc, and $M \leq -16.1$ at $D = 371$~Mpc. 

Several publications \citep[for example,][]{Nissanke2013, Singer2016, Gehrels2016} advocate that galaxy-targeted follow-up of GW triggers can be very effective when the event occurs within tens of megaparsecs. 
The discovery of AT2017gfo (the optical counterpart to GW170817) using a galaxy-targeted strategy is an example of success of this approach at a distance of 41~Mpc \citep{Coulter2017}.  However, at distances beyond $\sim 200$~Mpc, galaxy-targeted searches become more challenging.  
\cite{Gomez2019S190814bv} used the Magellan telescope to observe galaxies possibly hosting the S190814bv merger. In their work, \cite{Gomez2019S190814bv} imaged 96 galaxies at $3\sigma$ magnitude limit $i < 22.2$, corresponding to $M_i = -14.9$ mag at 267~Mpc. The sample includes all galaxies in the GLADE catalog within the 50\% probability volume with luminosity $\geq 0.15 L{^{*}}$.

\begin{figure}[t]
    \centering
        \includegraphics[width=1.\columnwidth]{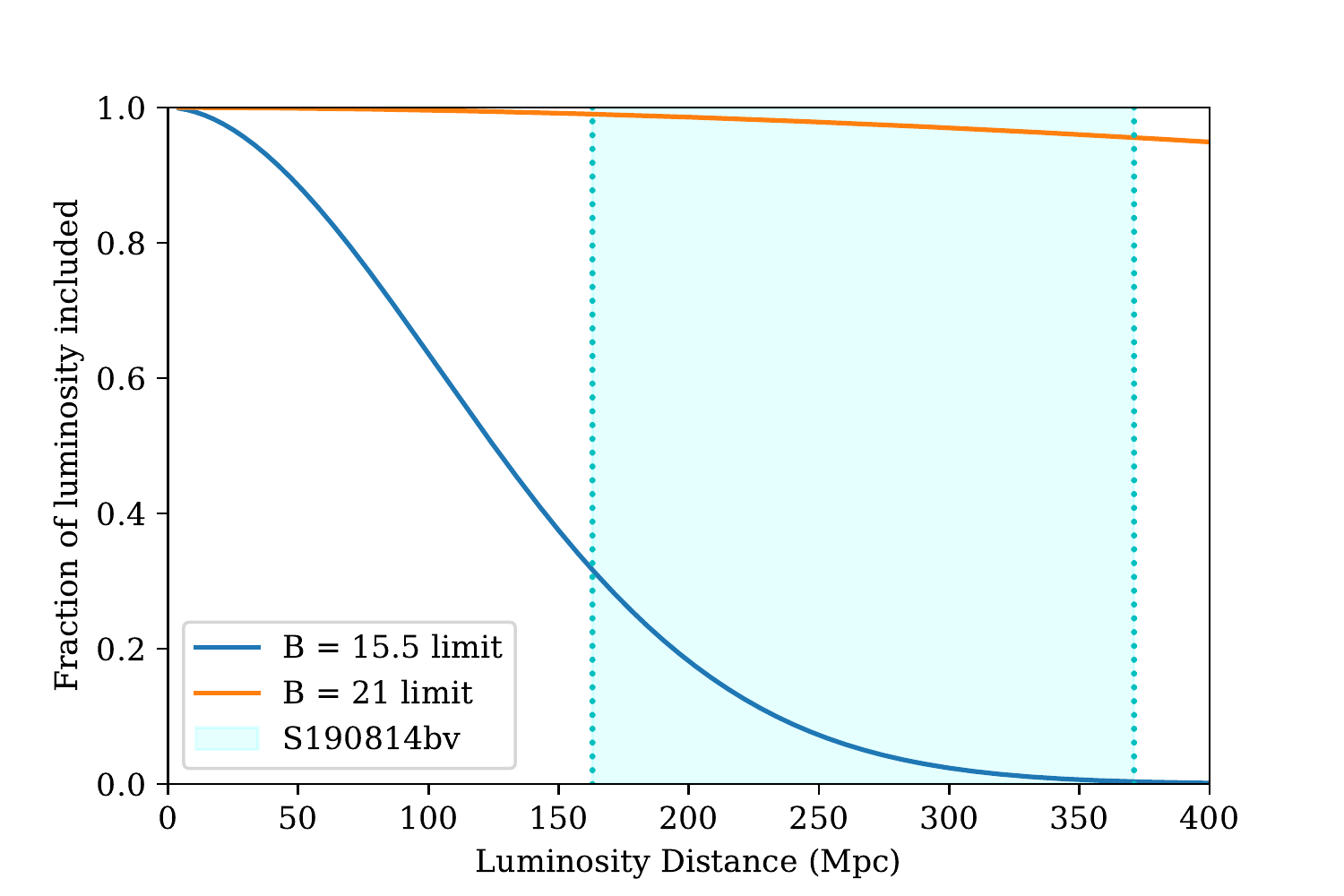}
        \includegraphics[width=1.\columnwidth]{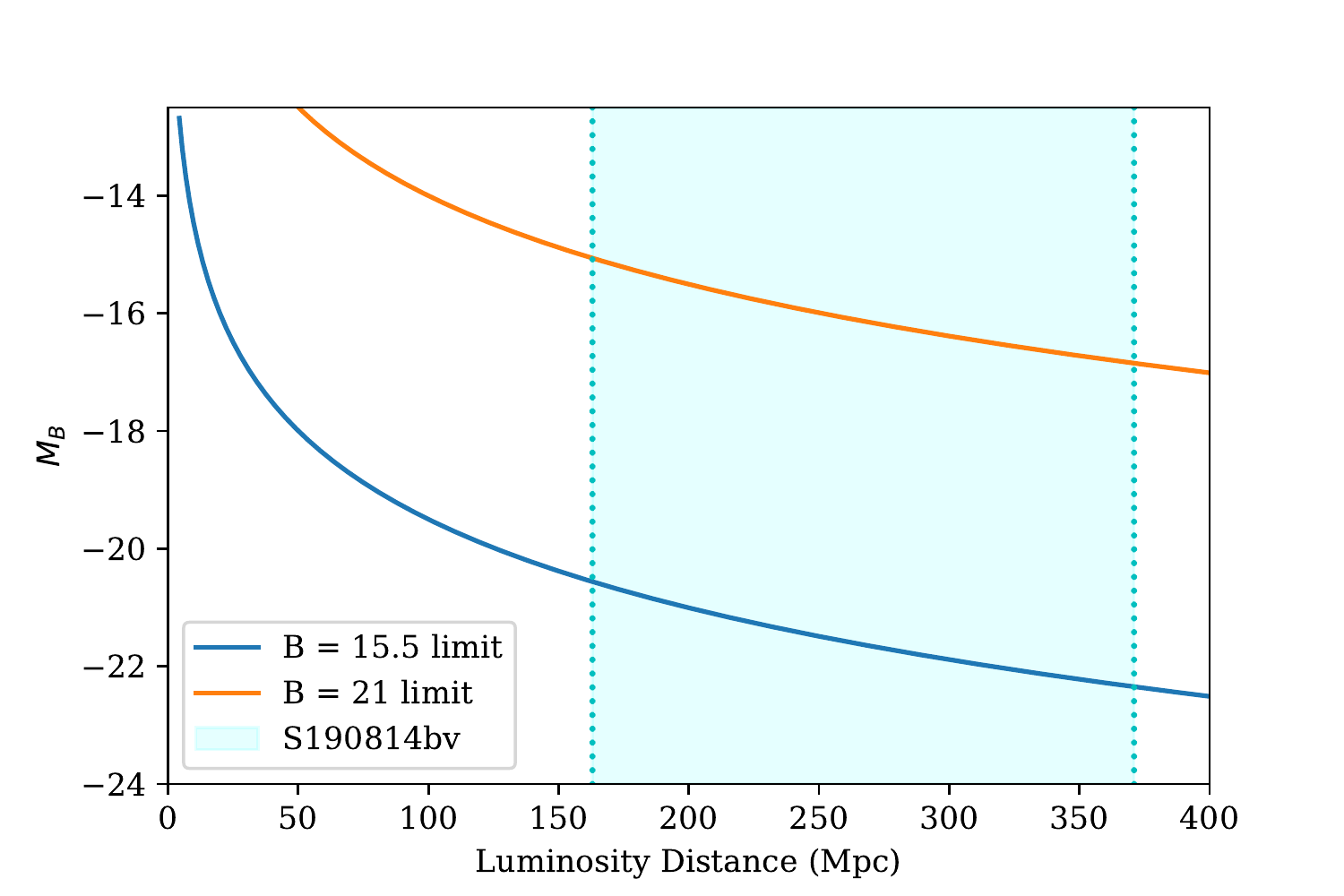}
    \caption{Luminosity fraction ({\it upper panel}) and $B$-band absolute magnitude limit ({\it lower panel}) as a function of luminosity distance. These quantities were  estimated for the Legacy Surveys DR8 photometric redshifts (orange line) as well as for the Galaxy List for the Advanced Detector Era \citep[GLADE,][blue line]{Dalya2018}
    assuming the catalog to be complete for $B < 21$ and $B < 15.5$ \citep{deVaucouleurs1991, Corwin1994} respectively, unsig a Schechter function as in \cite{Gehrels2016}.
    The 2$\sigma$ distance interval for S190814bv \citep{gcnS190814bvUpdate} is delimited by the cyan-colored dashed lines.  
    }
    \label{fig: galaxies}
\end{figure}

The analysis presented in this paper took advantage of photometric redshifts calculated from Legacy Surveys and {\it WISE} photometry mainly to exclude from our sample those candidates likely associated with galaxies significantly outside the distance range of S190814bv. Astrophysical transients with no clear association to a host galaxy were not excluded {\it a priori}, but their photometric evolution was not rapid enough for them to be considered likely counterparts to S190814bv. Nevertheless, we estimate the completeness that we could reach considering only a sample of transients found in the proximity of galaxies present in the photometric redshift catalog. Assuming a conservative limit of $B = 21$, we obtain a completeness $>97\%$ based on the luminosity fraction given a Schechter luminosity function \citep{Gehrels2016} in the $2\sigma$ distance range of S190814bv (Figure~\ref{fig: galaxies}). 
Although it is likely that the $z < 21$ excludes a large number of small, faint galaxies with $z > 21$, we are still nearly complete in luminosity. 
We note that $z$-band luminosity is a much better proxy for stellar mass than luminosity in bluer bands such as $B$, such that the spread in $z$ band mass-to-light ratio is smaller than the range of $B$/$z$ flux ratios amongst galaxies. As a result, the stellar mass completeness of the $z<21$ subset of DECaLS would be expected to be at least as high as the conservative $B$ luminosity completeness estimated here.

\section{Discussion}
\label{sec: discussion}

The results presented in Section~\ref{sec: results} show that no viable counterpart to S190814bv was discovered. 
In this section we discuss the constraints that this non-detection places on the astrophysical properties of the merger if the candidate was originally a neutron star--black hole.  

\subsection{Kilonova models}
\label{subsec: macronova models}

We used the upper limits obtained with DECam and kilonova simulations to constrain the parameter space of the possible EM counterpart to S190814bv. Specifically, we consider the kilonova models developed by \cite{Bulla2019} and \cite{Hotokezaka2019}.

We first compare DECam limits to 2D kilonova models computed with the Monte Carlo radiative transfer code \textsc{possis} \citep{Bulla2019}. These models assume a two-component ejecta geometry, with a lanthanide-rich component distributed around the equatorial plane and characterized by an half-opening angle $\phi$ and a lanthanide-poor component at higher latitudes (see Figure 1 in \citealt{Bulla2019NatAs}). Radiative transfer calculations are then performed to predict spectral time series for 11 different viewing angles, from which broad-band light-curves can be easily extracted. For our analysis, we choose $\phi=15^\circ$ and $\phi=30^\circ$ guided by numerical simulations \citep{Kawaguchi:2016ana,Fernandez2017} and calculate light curves for ejecta masses $M_\mathrm{ej}$ between 0.01 and $0.10$~$M_\odot$ (step size 0.01~$M_\odot$).

\begin{figure*}[t]
    \centering
        \includegraphics[width=2.\columnwidth]{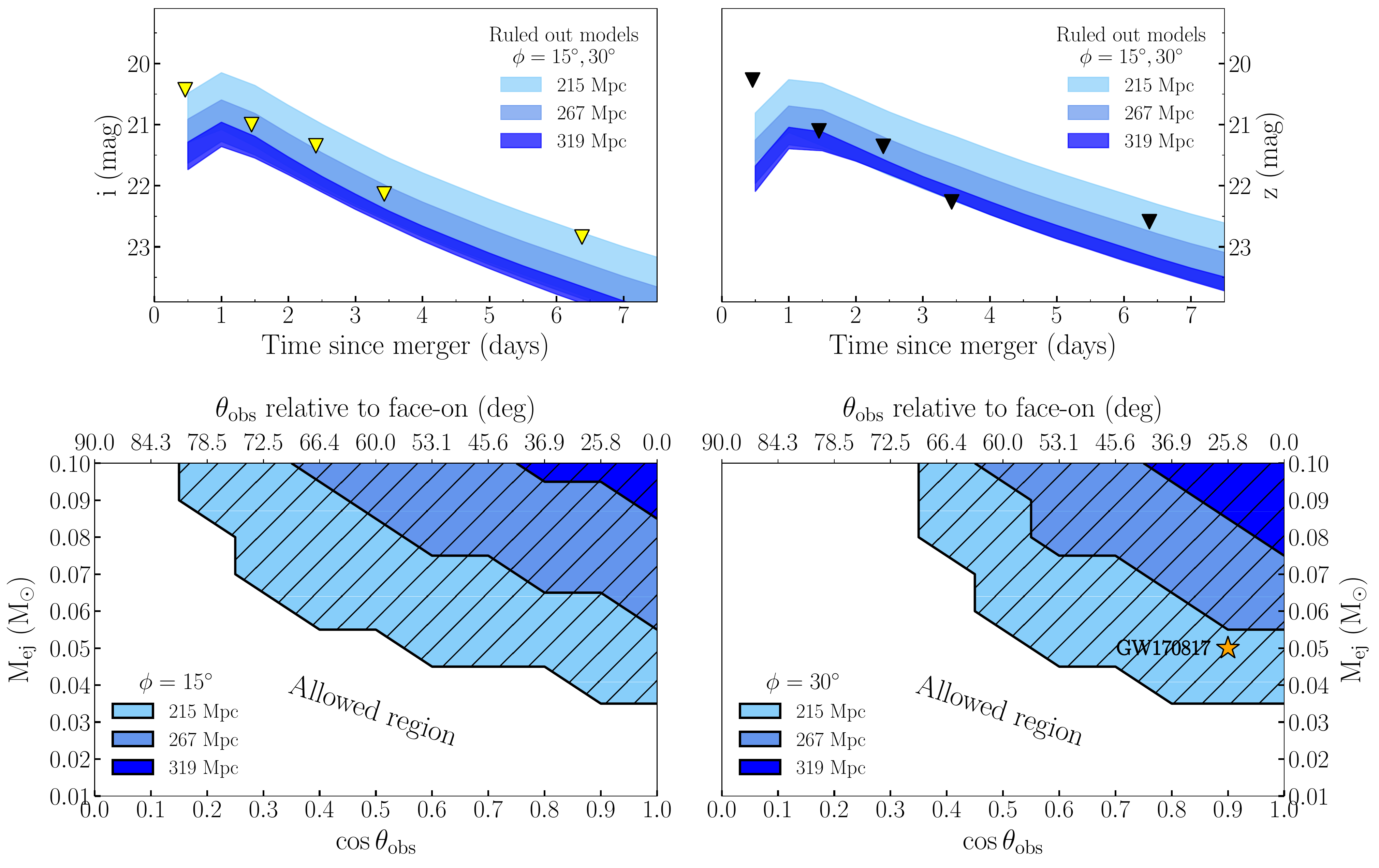}
    \caption{The limits obtained with DECam observations excluded regions of the parameter space for given kilonova models.  In this figure we consider models obtained with the Monte Carlo radiative transfer code \textsc{possis} \citep{Bulla2019} whose key parameters are the viewing angle $\theta_\mathrm{obs}$, the half-opening angle of an equatorial lanthanide-rich component $\phi$, and the ejecta mass $M_{ej}$. {\it Top --} $i$ (left) and $z$ (right) band light curves of kilonovae ruled out using the multi-band DECam upper limits (Table~\ref{table: limiting mags}), here marked with triangles. {\it Bottom --} Using the multi-band DECam upper limits, regions of the ejecta mass and viewing angle parameter space can be ruled out using $\phi=15^\circ$ (left) and $\phi=30^\circ$ (right). The best-fit model to GW170817 in this grid ($M_\mathrm{ej}=0.05$~$M_\odot$, $\phi=30^\circ$ and $\cos\theta_\mathrm{obs}=0.9$, \citealt{Dhawan2019}) is marked with a yellow star in the right panel. Both the {\it top} and {\it bottom} plots show that constraints on the models are more stringent if lower distances to S190814bv are considered. Here we used distances of 319~Mpc (dark blue patches), 267~Mpc (light blue patches), and 215~Mpc (cyan patches). }
    \label{fig:modelsBulla}
\end{figure*}

The top panels of Figure~\ref{fig:modelsBulla} show which modelled light curves are ruled out by DECam $i-$band (left) and $z-$band (right) limits for different distance assumptions (215, 267 and 319 Mpc from light to dark blue). As expected, more models are brighter than the limits and thus ruled out at closer compared to farther distances. Interestingly, we find that the most constraining limit is the $z-$band point at 3.4~days ($z$=22.3~mag), with all the other limits bringing no improvement in terms of ruling out models. We note that comparable deep limits at earlier epochs, when the kilonova is intrinsically brighter, would have been extremely important to constrain the parameter space more strongly. 

The bottom panels of Figure~\ref{fig:modelsBulla} show what region of the $M_\mathrm{ej}$ - viewing angle parameter space is ruled out for $\phi=15^\circ$ (left) and $\phi=30^\circ$ (right). The brightest kilonovae in the modelled grid are predicted at high $M_\mathrm{ej}$ and for polar viewing angles (system viewed face-on, $\theta_\mathrm{obs}=0$ and $\cos\theta_\mathrm{obs}=1$). These models are therefore the first to be ruled out by DECam limits (upper-right corner in the $M_\mathrm{ej}$ - viewing angle parameter space). Stronger constraints are found for closer distances (see above) and smaller $\phi$~angles as the larger contribution of the lanthanide-poor compared to lanthanide-rich component leads to an intrinsically brighter kilonova. We note that the best-fit model to GW170817 in this grid ($M_\mathrm{ej}=0.05$~$M_\odot$, $\phi=30^\circ$ and $\cos\theta_\mathrm{obs}=0.9$, \citealt{Dhawan2019}) would be slightly fainter and thus hidden below DECam limits at 267~Mpc.
To summarise, ejecta masses are constrained to $M_\mathrm{ej}<0.04$~$M_\odot$ in the most optimistic case assuming the nearest consistent distance of 215 Mpc, $\phi=15^\circ$ and $\cos\theta_\mathrm{obs}=1$ (face-on). A more conservative constraint ($M_\mathrm{ej}\lesssim0.10$~$M_\odot$) is instead found for farther distances, viewing angles closer to the equatorial plane and larger $\phi$ values.

\begin{figure}[t]
    \centering
        \includegraphics[width=1.\columnwidth]{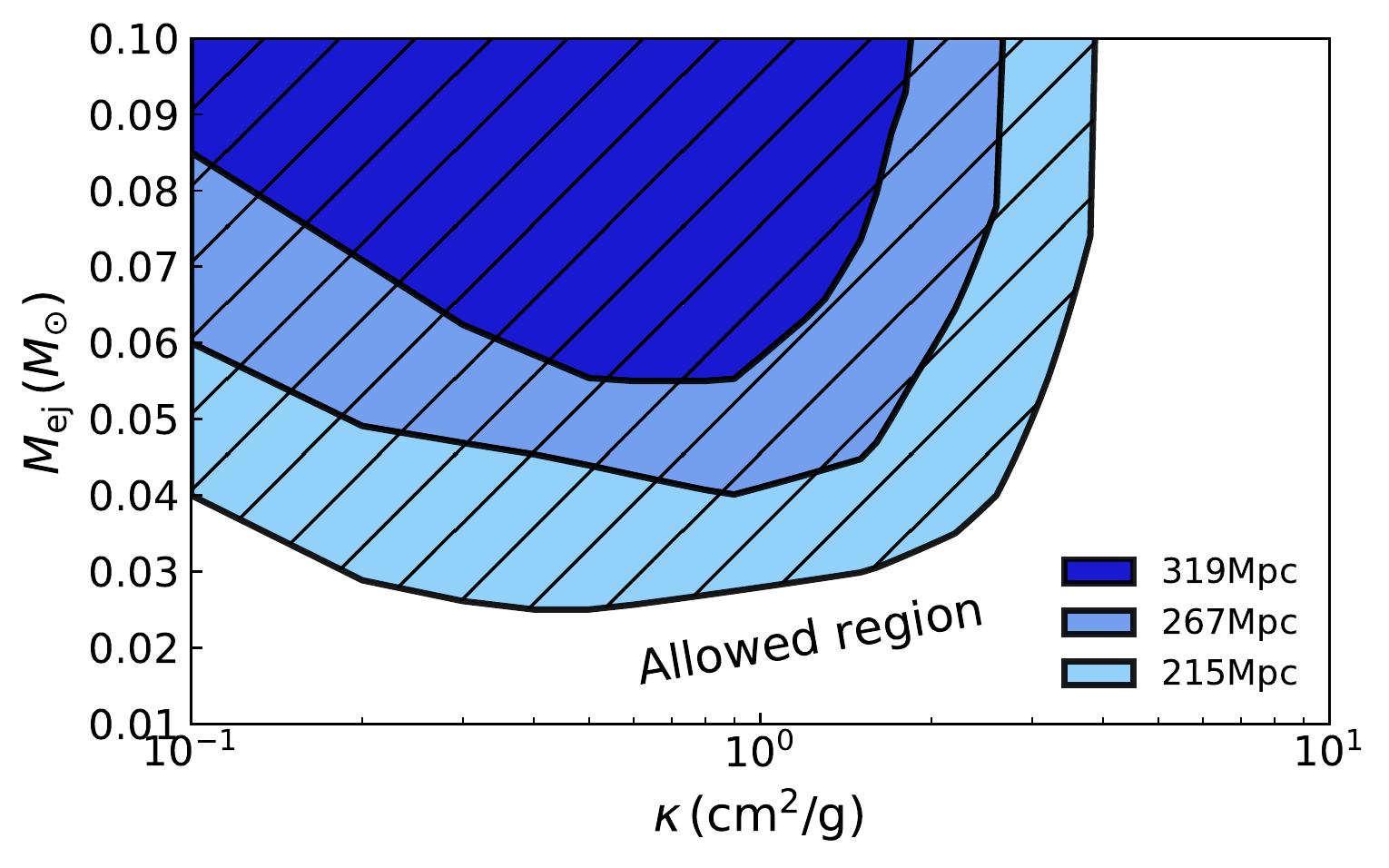}
    \caption{Constraints on the ejecta mass ($M_{ej}$) and opacity ($\kappa$) phase space obtained using multi-band DECam upper limits (Table~\ref{table: limiting mags}) and the kilonova models described in \cite{Hotokezaka2019}. Similarly to Figure~\ref{fig:modelsBulla}, the constraints become more significant assuming lower distances to the merger. }
    \label{fig: models Hotokezaka}
\end{figure}

Figure \ref{fig: models Hotokezaka} presents upper limits on the ejecta mass obtained using a different approach. We assume a spherical ejecta with a power-law density profile $\rho \propto v^{-n}$ for $v_{min}<v<v_{max}$ and calculate the emission using the heating rate formalism and light curve modeling described in \cite{Hotokezaka2019}.
The outflow parameters are $v_{min}=0.1$c, $v_{max}=0.4$c and $n=4.5$. The composition that we consider is of r-process elements with atomic mass $85 \leq A \leq 209$ and a solar abundance pattern. The heating-rate calculation includes only $\beta$-decay. We assume further that the entire ejecta can be characterised by a single grey opacity parameter $\kappa$ and vary the value of $\kappa$. The shaded regions in the $M_{ej}$-$\kappa$ space in Figure~\ref{fig: models Hotokezaka} are where the light curve is brighter than the upper limits we have for this event. The conclusion from this figure is that the ejecta cannot have more than $\sim 0.05~M_\odot$ of ejecta that is not lanthanide rich at a distance of 267~Mpc, or $\sim 0.03~M_\odot$ at an optimistic distance of 215~Mpc.  This conclusion is in agreement with the results obtained with the \cite{Bulla2019} kilonova models under favorable ($\theta \lesssim 30^{\circ}$) viewing angles. 

\subsection{Constraints on the merging binary}
\label{subsec: binary}

At present, constraints on the amount of mass ejected by the merger can be translated into approximate constraints on the initial parameters of the possible merging neutron star and black hole (see \citealt{Coughlin:2019zqi} for a summary of other events during O3a). As a proof of principle, given the mass ratio of the binary $Q=M_{\mathrm{BH}}/M_{\rm NS}$, the dimensionless component of the initial black hole spin aligned with the orbital angular momentum ($\chi_{\rm aligned}$), and the compactness of the neutron star $C_{\rm NS}=GM_{\rm NS}/(R_{\rm NS}c^2)$, we can conservatively assume that:
\begin{equation}
M_{\rm ej} \gtrsim M_{\rm dyn} + 0.1 (M_{\rm out}-M_{\rm dyn}),
\nonumber
\end{equation}
where $M_{\rm out}$ represents the mass that remains outside of the black hole after merger~\citep{Foucart:2012nc,Foucart:2018rjc}, and $M_{\rm dyn}$ denotes the mass ejected during disruption~\citep{Kawaguchi:2016ana}. Both $M_{\rm out}$ and $M_{\rm dyn}$ are predictions from semi-analytical fits to the results of merger simulations. In the above, we have also conservatively assumed that $\gtrsim 10\%$ of the matter initially bound in an accretion disk around the remnant black hole will be ejected in magnetically-driven and/or neutrino-driven winds, and during viscous expansion of the disk~\citep{Fernandez:2013tya,Fernandez:2014cna,Siegel:2017nub,Christie2019MNRAS}.\footnote{For low mass black holes leaving remnants comparable to the initial conditions of existing 3D simulations, $\gtrsim 25\%$ of the disk is most likely ejected, but more compact disks around massive black holes eject a smaller fraction of their disk.}

\begin{figure}[t]
    \centering
        \includegraphics[width=1.\columnwidth]{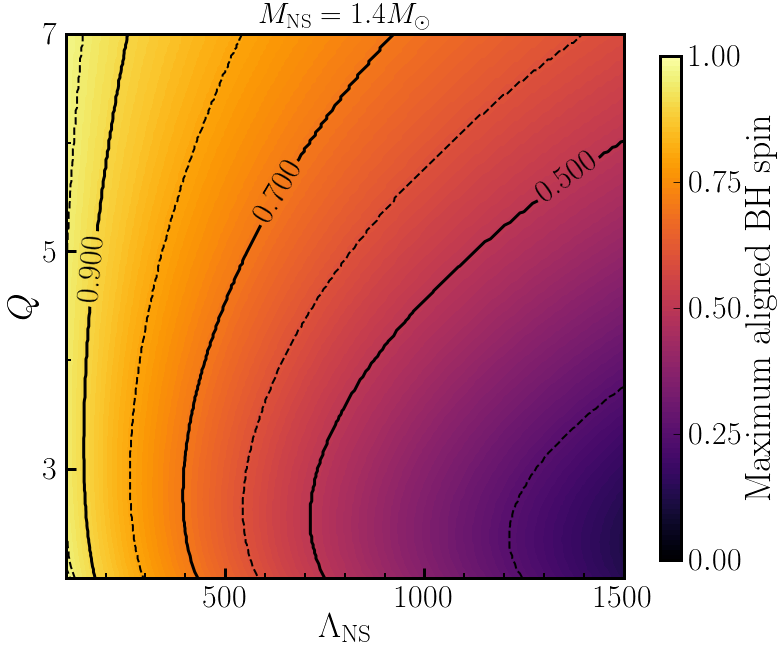}
    \caption{Constraints on the parameter space of black hole-neutron star binaries assuming M$_{ej} < 0.03$~M$_\odot$. We show the highest possible value of the component of the black hole spin aligned with the orbital momentum as a function of mass ratio $Q=M_{\rm BH}/M_{\rm NS}$ and tidal deformability $\Lambda_{\rm NS}$ of the neutron star. The figure obtained assumes that $M_{\rm NS}=1.4M_\odot$, but as the ejected mass is approximately proportional to $M_{\rm NS}$ (at fixed $\Lambda_{\rm NS}$), any choice in the range $M_{\rm NS}\sim (1.2-1.6)M_\odot$ would give qualitatively similar constraints.}
    \label{fig:BHNSparams}
\end{figure}

In Figure~\ref{fig:BHNSparams}, we show constraints on the parameter space of NSBH binaries assuming $M_{\rm ej}=0.03M_\odot$.\footnote{We do not show the conservative case of $M_{\rm ej}<0.1M_\odot$ as it does not provide meaningful constraints on the parameter space of the binary because the fitting formulae are not properly calibrated above $\chi_{\rm aligned}=0.9$} Practically, an upper bound on $M_{\rm ej}$ can be interpreted as a maximum possible value of $\chi_{\rm aligned}$ for each choice of mass ratio and dimensionless neutron star deformability $(Q,\Lambda_{\rm NS})$, or an upper bound on $\Lambda_{\rm NS}$ at fixed $\chi_{\rm aligned}$. $\Lambda_{\rm NS}$ is related to the neutron star EOS, and is given by $\Lambda_\mathrm{NS} = \frac{\lambda c^{10}}{G^4 M_\mathrm{NS}^5}$, where $\lambda = 2k_2 R_\mathrm{NS}^5 / (3G)$ and  $k_2$ is the Love number \citep[see e.g.,][]{Flanagan2008,Hinderer2010}.

Assuming $\Lambda \sim 800$ \citep[the largest value allowed by GW170817 for a 1.4~M$_{\odot}$ NS,][]{Abbott2017GW170817discovery} and $M_{ej} < 0.03$~M$_{\odot}$, the data constrain the BH spin component aligned with the orbital momentum to be $\chi < 0.7$ for mass ratios $Q < 6$.
This constraint becomes looser for more compact stars (lower $\Lambda$), and tighter for less compact stars (larger $\Lambda$).

\section{Conclusion}
\label{sec: conclusion}

In this paper, we have presented deep synoptic limits on the optical counterpart to the NSBH merger S190814bv by analyzing publicly available data from a DECam imaging campaign. 
We identified dozens of counterpart candidates, and systematically ruled each of them out using the results of a global follow-up campaign undertaken by our group and the community. Real-time data analysis and prompt follow-up allowed the candidates to be classified at timescales from hours to days. 
Based on our lack of identification of an optical counterpart, we used our detection limits and kilonova models to constrain the allowable parameter space for S190814bv. 
We found that the ejecta mass can be poorly constrained at the far end of the distance probability distribution, however limits on the ejecta mass of $M_{ej} \lesssim 0.05$~M$_{\odot}$ can be placed at a luminosity distance of 267~Mpc at polar viewing angles or assuming an opacity $\kappa < 2$~cm$^2$g$^{-1}$.  A more stringent limit of $M_{ej} \lesssim 0.03$~M$_{\odot}$ can be placed assuming a distance of 215~Mpc. Using the constrains that we obtained for the ejecta mass, we showed how the phase space of the NSBH binary system can also be constrained. In particular, reliable constraints on the highest possible value of the BH spin component aligned with the orbital momentum as a function of $Q$ and $\Lambda_{NS}$ can be placed for $M_{ej} < 0.03$~M$_{\odot}$. For example, assuming a tidal deformability at the high end of the range allowed by gravitational wave observations of GW170817, we can constrain the spin component to be $\chi < 0.7$ for mass ratios $Q < 6$.
The non-detection of a viable counterpart to S190814bv is also consistent with the progenitor being a low-mass binary BH, rather than a NSBH system.

Follow-up observations of new NSBH mergers with DECam and transient characterization with telescope networks have great potential to unveil electromagnetic counterparts during O3 and beyond. Non-detections such as this one  can significantly constrain the parameter space of NSBH merger and kilonova models. In the near future, follow-up campaigns with LSST will allow us to probe NSBH mergers deeper and at larger distances. 

\acknowledgements
\section*{Acknowledgments}

This work was supported by the GROWTH (Global Relay of Observatories Watching Transients Happen) project funded by the National Science Foundation under PIRE Grant No 1545949. GROWTH is a collaborative project among California Institute of Technology (USA), University of Maryland College Park (USA), University of Wisconsin Milwaukee (USA), Texas Tech University (USA), San Diego State University (USA), University of Washington (USA), Los Alamos National Laboratory (USA), Tokyo Institute of Technology (Japan), National Central University (Taiwan), Indian Institute of Astrophysics (India), Indian Institute of Technology Bombay (India), Weizmann Institute of Science (Israel), The Oskar Klein Centre at Stockholm University (Sweden), Humboldt University (Germany), Liverpool John Moores University (UK) and University of Sydney (Australia).

Daniel A. Goldstein acknowledges support from Hubble Fellowship grant HST-HF2-51408.001-A.
Support for Program number HST-HF2-51408.001-A is provided by NASA through a grant from the Space Telescope Science Institute, which is operated by the Association of Universities for Research in Astronomy, Incorporated, under NASA contract NAS5-26555.
We gratefully acknowledge Amazon Web Services, Inc. for a generous grant (\texttt{PS\_IK\_FY2019\_Q3\_ Caltech\_Gravitational\_Wave}) that funded our use of the Amazon Web Services cloud computing infrastructure to process the DECam data.
P. E. Nugent acknowledges support from the DOE through DE-FOA-0001088, Analytical Modeling for Extreme-Scale Computing Environments.  D.~A. Perley and D.~A.~Goldstein performed the work associated with this project at the Aspen Center for Physics which is supported by National Science Foundation grant PHY-1607611.  This work was partially supported by a grant from the Simons Foundation.
AJCT thanks I. Agudo, J. Cepa, V. Dhillon, J. A. Font, A. Martin-Carrillo, S. R. Oates, S. B. Pandey, E. Pian, R. Sanchez-Ramirez, A. M. Sintes, V. Sokolov and B.-B. Zhang for fruitful conversations.
F. Foucart gratefully acknowledges support from NASA through grant 80NSSC18K0565 and from the NSF through grant PHY1806278.
M. Bulla, A. Goobar, E. Kool, S. Dhawan, and J. Sollerman acknowledge support from the G.R.E.A.T research environment funded by the Swedish National Science Foundation.
 J.~Sollerman acknowledges support from the Knut and Alice Wallenberg Foundation.
 J.\,S.~Bloom and K. Zhang are partially supported by a Gordon and Betty Moore Foundation Data-Driven Discovery grant. 
 D.~A.~H. Buckley acknowledges research support from the National Research Foundation of South Africa.
 M.~W.~Coughlin is supported by the David and Ellen Lee Postdoctoral Fellowship at the California Institute of Technology.
 S.~Nissanke and G.~Raaijmakers are grateful for support from VIDI, Projectruimte and TOP Grants of the Innovational Research Incentives Scheme (Vernieuwingsimpuls) financed by the Netherlands Organization for Scientific Research (NWO).
H. Kumar and K. Zhang thank the LSSTC Data Science Fellowship Program, which is funded by LSSTC, NSF Cybertraining Grant \#1829740, the Brinson Foundation, and the Moore Foundation; his participation in the program has benefited this work.
 D.~Dobie is supported by an Australian Government Research Training Program Scholarship.
 P. Gatkine is supported by NASA Earth and Space Science Fellowship (ASTRO18F-0085).
D.~L.~Kaplan was supported by NSF grant AST-1816492.
Y.D. Hu thanks the support by the program of China Scholarships Council (CSC) under the Grant No.201406660015.
A.~K.~H.~Kong acknowledges support from the Ministry of Science and Technology of the Republic of China (Taiwan) through grants 107-2628-M-007-003 and 108-2628-M-007-005-RSP.
V. Z. Golkhou acknowledges support from the University of Washington College of Arts and Sciences, Department of Astronomy, and the DIRAC Institute. University of Washington's DIRAC Institute is supported through generous gifts from the Charles and Lisa Simonyi Fund for Arts and Sciences, and the Washington Research Foundation. M.~Juric and A.~Connolly acknowledge the support of the Washington Research Foundation Data Science Term Chair fund, and the UW Provost's Initiative in Data-Intensive Discovery.
S. Mohite thanks the LSSTC Data Science Fellowship Program, which is funded by LSSTC, NSF Cybertraining Grant-1829740, the Brinson Foundation, and the Moore Foundation; his participation in the program has benefited this work.
MG is supported by the Polish NCN MAESTRO grant 2014/14/A/ST9/00121. 

This research has made use of the VizieR catalogue access tool, CDS, Strasbourg, France (DOI: \texttt{10.26093/cds/vizier}). The original description of the VizieR service was published in A\&AS 143, 23.
This project used data obtained with the Dark Energy Camera (DECam), which was constructed by the Dark Energy Survey (DES) collaborating institutions: Argonne National Lab, University of California Santa Cruz, University of Cambridge, Centro de Investigaciones Energeticas, Medioambientales y Tecnologicas-Madrid, University of Chicago, University College London, DES-Brazil consortium, University of Edinburgh, ETH-Zurich, University of Illinois at Urbana-Champaign, Institut de Ciencies de l'Espai, Institut de Fisica d'Altes Energies, Lawrence Berkeley National Lab, Ludwig-Maximilians Universitat, University of Michigan, National Optical Astronomy Observatory, University of Nottingham, Ohio State University, University of Pennsylvania, University of Portsmouth, SLAC National Lab, Stanford University, University of Sussex, and Texas A$\&$M University. Funding for DES, including DECam, has been provided by the U.S. Department of Energy, National Science Foundation, Ministry of Education and Science (Spain), Science and Technology Facilities Council (UK), Higher Education Funding Council (England), National Center for Supercomputing Applications, Kavli Institute for Cosmological Physics, Financiadora de Estudos e Projetos, Funda\c{c}\~{a}o Carlos Chagas Filho de Amparo a Pesquisa, Conselho Nacional de Desenvolvimento Cient\'{i}fico e Tecnol\'{o}gico and the Minist\'{e}rio da Ci\^{e}ncia e Tecnologia (Brazil), the German Research Foundation-sponsored cluster of excellence ``Origin and Structure of the Universe" and the DES collaborating institutions.

\noindent The Liverpool Telescope is operated on the island of La Palma by Liverpool John Moores University in the Spanish Observatorio del Roque de los Muchachos of the Instituto de Astrofisica de Canarias with financial support from the UK Science and Technology Facilities Council.
\noindent Based on observations made with the Gran Telescopio Canarias (GTC),
installed in the Spanish Observatorio del Roque de los Muchachos of the
Instituto de Astrof\'isica de Canarias, in the island of La Palma. This
work is partly based on data obtained with the instrument OSIRIS, built
by a Consortium led by the Instituto de Astrof\'isica de Canarias in
collaboration with the Instituto de Astronom\'ia of the Universidad
Aut\'onoma de M\'exico. OSIRIS was funded by GRANTECAN and the National Plan
of Astronomy and Astrophysics of the Spanish Government.
Some of the observations reported in this paper were obtained with the Southern African Large Telescope (SALT). Polish participation in SALT is funded by grant no. MNiSW DIR/WK/2016/07.

\bibliography{references,gcn_S190814bv}
\bibliographystyle{apj}

\end{document}